\documentclass[12pt]{article}
\usepackage{epsfig}
\usepackage{color}
\usepackage{amssymb,amsmath}
\usepackage{graphicx}
\usepackage{epsfig}

\newcommand{\bea}{\begin{eqnarray}}
\newcommand{\eea}{\end{eqnarray}}
\newcommand{\met}{$E_{{\rm T}}$\hspace{-1.10em}/  \ \ }

 \makeatletter
\def\alt{\mathrel{\mathpalette\gl@align<}}
\def\agt{\mathrel{\mathpalette\gl@align>}}
\def\gl@align#1#2{\lower.6ex\vbox{\baselineskip\z@skip\lineskip\z@
\ialign{$\m@th#1\hfil##\hfil$\crcr#2\crcr\sim\crcr}}} \makeatother

\begin{document}
\begin{flushright}
KEK-TH-1577 \\
UT-12-21
\end{flushright}
\vspace*{1.0cm}

\begin{center}
\baselineskip 20pt 
{\Large\bf 
Isospin-Violating Dark Matter at the LHC
}
\vspace{1cm}

{\large 
Kaoru Hagiwara$^{a}$, Danny Marfatia$^{b}$, Toshifumi Yamada$^{a,c}$
} \vspace{.5cm}

{\baselineskip 20pt \it

$^{a}$ KEK Theory Center and SOKENDAI, \\
1-1 Oho, Tsukuba, Ibaraki 305-0801, Japan \\ \
\\

$^{b}$ Department of Physics \& Astronomy, University of Hawaii, Honolulu, HI 96822, U.S.A
\\
and\\
Department of Physics \& Astronomy, University of Kansas, 
Lawrence, KS 66045, USA \\ \
\\
$^{c}$ Department of Physics, University of Tokyo, \\
7-3-1 Hongo, Bunkyo-ku, Tokyo 113-0033, Japan }

\vspace{.5cm}

\vspace{1.5cm} {\bf Abstract} \end{center}

\ \ \ We consider a toy model of dark matter (DM)
 with a gauge singlet Dirac fermion that 
 has contact interactions to quarks that differ
 for right-handed up and down quarks.
This is motivated by the isospin-violating dark matter scenario that was proposed to
 reconcile reported hints of direct DM detection with bounds from non-observation of the signal
 in other experiments.
We discuss how the effects of isospin violation in these couplings can be observed at the LHC.
By studying events with large missing transverse momentum (\met),
 we show that the ratio of mono-photon and mono-jet events is sensitive to
 the ratio of the absolute values of the couplings to the up and down quarks,
 while a dedicated study of di-jet plus \met events can reveal their relative sign. We also
 consider how our results are modified if instead of a contact interaction, a particle that mediates the interaction
 is introduced.
 Our methods have broad applicability to new physics that involves
 unequal couplings to up and down quarks.

\thispagestyle{empty}

\newpage

\setcounter{footnote}{0}
\baselineskip 18pt
%

\section{Introduction}

\ \ \ Once a hint of dark matter (DM) is found in events with large missing transverse 
 momentum (\met) at the LHC,
 the next task will be to study how the DM particle couples to standard model (SM) particles.
It can be the case that the DM particle couples to up and down quarks differently.
Observing such an isospin violating feature of the new physics sector
would give an important clue for the determination of the Lagrangian of the underlying theory.
Also, it is itself a challenge in the physics of hadron colliders,
 requiring elaborate kinematical cuts to extract information from parton-level processes.

In this context,  
 isospin-violating dark matter (IVDM)~\cite{ivd} provides a scenario in which
a DM particle does not couple identically to up and down quarks.
This model was motivated by DM direct detection experiments.
The DAMA~\cite{dama} and CoGeNT~\cite{cogent} experiments
 observed signals that are potentially of dark matter origin. The signals 
 are consistent with a DM particle of mass 10~GeV 
 scattering off nuclei with a spin-independent nucleon cross section of 
 $\sigma_{N} \sim 2 \times 10^{-4}$~pb and $\sigma_{N} \sim 7 \times 10^{-5}$~pb,
 respectively.
However, the XENON~\cite{xenon} and CDMS~\cite{cdms} experiments reported negative results.
To accommodate the apparently inconsistent data,
 the authors of Ref.~\cite{ivd} considered a dark matter particle
 that couples to up and down quarks differently.
Then the cross section of dark matter scattering off protons differs from that off neutrons,
 and the apparent tension among dark matter direct detection experiments
 is attributed to the different proton-neutron ratios of the detector materials.
Recently, the LUX experiment~\cite{lux} has reported a stringent upper bound on the spin-independent cross section 
 for DM elastic scattering off nuclei.
A tension between the LUX and the DAMA/CoGeNT results persists even if isospin-violating couplings of DM are assumed~\cite{gresham}, because liquid xenon
 is composed of several isotopes so that it is not possible to completely suppress xenon's coupling to DM~\cite{ivd}.
More recently, the SuperCDMS~\cite{supercdms} and CDEX~\cite{cdex} experiments have also reported null results that corroborate the LUX results.
Nevertheless, independently of the aforementioned anomalies in the context of light DM, the possibility that a generic DM particle interacts with SM quarks through isospin-violating couplings
 remains and should be investigated.

We study the possibility of testing the scenario at the LHC
 by measuring the couplings to up and down quarks.
We work in a general context and simply assume that
 a new stable particle that is a singlet under the SM gauge group
 (called ``dark matter" in the following) couples to right-handed up and down quarks
 with different strengths and signs.
At the LHC, the production of DM particles associated with a jet(s) or photon
 gives rise to events with a hard jet(s) or photon plus large missing transverse momentum (\met).
Since the up and down quarks couple differently to the photon but identically to the gluons,
 the magnitudes of the two couplings can be measured
 by comparing the cross section of the mono-jet plus \met signal
 with that of the mono-photon plus \met signal.
The relative sign of the dark matter couplings is more difficult to measure
 because it requires the measurement of the interference between
 the dark matter coupling to the up quark and that to the down quark.
This is possible through the subprocess, $ud \rightarrow ud + DM \, DM$,
 but the interference effects need to be identified in the presence of the other dijet$ + DM \, DM$
 subprocesses and the SM backgrounds.

Our discussion of how isospin-violating couplings of a DM particle
 can be studied at the LHC involves several ingredients.
We use MadGraph~5~\cite{mg} for calculating matrix elements
 and for generating DM signal events,
 Pythia~8~\cite{pythia} for parton showering, and PGS~4~\cite{pgs} for simulating detector effects.
The outline of the paper is as follows. In Section~2, we describe our toy model of IVDM.
In Sections~3, 4 and 5, we investigate the DM signals from three channels,
 mono-jet plus \met, mono-photon plus \met, and di-jet plus \met.
In Section~6, we present the latest LHC bounds on the contact interaction scale for couplings that
suppress scattering in xenon detectors.
In Section~7, we consider a model in which isospin-violating interactions result from a mediator field that couples to the DM
 and quarks, and study how the presence of the mediator enhances the DM signal cross sections.
Section~8 is devoted to discussion and conclusion.
\\

\section{Model}

\ \ \ We consider a toy model in which a DM particle couples to quarks
 with isospin-violating couplings.
We introduce a Dirac fermion $\chi$ that is a singlet under the SM gauge group.
It couples to the up and down quarks through vector-like and axial vector-like contact interactions.
The relevant part of the Lagrangian is
\begin{eqnarray}
{\cal L}_{DM} &=& i \bar{\chi} \gamma^{\mu} \partial_{\mu} \chi - m_{\chi} \bar{\chi} \chi
%
+ \frac{1}{\Lambda^{2}}  
( g_{Q} \bar{q}_{L} \gamma_{\mu} q_{L} \ + \ g_{U} \bar{u}_{R} \gamma_{\mu} u_{R} 
\ + \ g_{D} \bar{d}_{R} \gamma_{\mu} d_{R} ) (\bar{\chi} \gamma^{\mu} \chi) \ ,
\label{lag}
\end{eqnarray}
 where $\Lambda$ denotes the energy scale of the contact interaction and $q_{L}$ denotes the SU(2) doublet
 of left-handed up and down quarks.
For $g_{U}\neq g_{D}$, $\chi$ is IVDM.

At the LHC, we would like to extract the magnitudes and relative signs of the couplings, $g_{U}$, $g_{D}$
 and $g_{Q}$.
In the following, we examine the possibility of measuring $g_{U}/g_{D}$ by setting $g_{Q}=0$.
Implications of $g_{Q}\neq 0$ will be discussed in the last section.
\\

The (absolute) perturbative unitarity bounds for the DM production processes 
 $u_R \bar{u}_R \rightarrow \chi \bar{\chi}$ and $d_R \bar{d}_R \rightarrow \chi \bar{\chi}$
 are saturated when the center-of-mass energy, $s$, satisfies
\begin{eqnarray}
\sqrt{s} &=& (48 \pi^2)^{1/4} \ \frac{\Lambda}{\sqrt{g_q}} \ \ \ (q=u,d) \ . \label{uni bound for lag}
\end{eqnarray}
For example, for $\Lambda=800$ GeV and $g_q=1$, 
 the perturbative unitarity bound is violated for $\sqrt{s} > 3.7$ TeV.
Throughout, we simply assume
 the cross section for a DM production process to be constant for
 $\sqrt{s} \geq (48 \pi^2)^{1/4} \ \Lambda/\sqrt{g_q}$,
 although it will turn out that the results of our analysis are insensitive to
 the behavior of DM production cross sections for $\sqrt s$ near and above the unitarity bound.
A detailed discussion of unitarity bounds for inelastic scattering
 is provided in the Appendix.
\\

\section{Mono-jet + \met}

\ \ \ In this section, we study events with a high $p_{T}$ mono-jet and large \met,
 which is the discovery channel for a DM particle at hadron colliders.
We first focus on the case of $pp$ collisions at $\sqrt{s}=7$ TeV
 and reproduce the results of Ref.~\cite{fox}.
We then consider $\sqrt{s}=14$ TeV, the target energy of the LHC.
By optimizing the selection cuts, we estimate the DM signal for various values of $g_{U}/g_{D}$
 for a fixed value of $\sqrt{g_{U}^{2} + g_{D}^{2} }$ and $g_{Q}=0$.

\subsection{Study for $\sqrt{s}=7$ TeV}

\ \ \ Following Ref.~\cite{fox}, we adopt the \textit{veryHighPT} cut of the ATLAS search~\cite{atlas},
 which yields the most stringent bound on the DM-quark interactions.
This cut is described by

\begin{itemize}
 \item Require \met $>  300$~GeV.
 \item Highest $p_{T}$ jet should satisfy $\vert \eta_{1} \vert < 2.0$ and $p_{T 1} > 350$~GeV.
 \item Event is vetoed if the second hardest jet satisfies $\vert \eta_{2} \vert < 4.5$, and 
 $p_{T 2} > 60$~GeV or 
 $\Delta \phi (\vec{p}_{T 2}, \vec{E}_{{\rm T}}\hspace{-1.10em}/ \ \ \ ) < 0.5$.
 \item Event is vetoed if there is another jet that satisfies $\vert \eta \vert < 4.5$ and $p_{T} > 30$~GeV.
 \item Event is vetoed if there is an electron that satisfies $\vert \eta_{e} \vert < 2.47$ and $p_{T e} > 20$~GeV.
 \item Event is vetoed if there is a muon that satisfies $\vert \eta_{\mu} \vert < 2.4$ and $p_{T \mu} > 10$~GeV.
\end{itemize}

The dominant sources of background for mono-jet events that satisfy the above criteria
 are $Z$ + jet events with $Z \rightarrow \nu \nu$, and 
 $W$ + jet events with $W \rightarrow \tau \nu$
 or $W \rightarrow l \, \nu \ (l=e, \mu)$ where the $\tau$ decay products or leptons
 do not satisfy any of the veto criteria.

We estimate the $Z$ + jet background by first generating
 $Z (\rightarrow \nu \nu)$ + 1, 2 jets events
 at the matrix element level \cite{mg} with precuts \met$ > 200$ GeV and $k_{T} > 140$ GeV for the jets.
We next simulate parton showering \cite{pythia}
 by using the $k_{T}$-jet matching scheme \cite{mlm}
 to match the $p_{T}$ distribution of the second hardest jet
 simulated via the parton showering of $Z$ + 1 jet events
 and the distribution calculated from the matrix elements of $Z$ + 2 jets events.
The matching scale is set at 200~GeV.
We have confirmed that the cross section after the final cut
 does not change significantly when the matching scale is varied by $\pm 60$~GeV around 200~GeV.
Finally, we perform a detector simulation~\cite{pgs} and 
 the cross section for $Z (\rightarrow \nu \nu)$ + jets events that pass the \textit{veryHighPT} cut
 is found to be
\begin{eqnarray}
\sigma_{Z(\rightarrow \nu \nu)+{\rm jets}} \ (veryHighPT) &\simeq& 106 \ {\rm fb}\,.
\label{eq2}
\end{eqnarray}
In a similar manner, we generate
 $W$ + 1, 2 jets events in which $W$ decays into $\tau \nu_{\tau}$
 and those in which $W$ decays into $e \nu_{e}$ or $\mu \nu_{\mu}$,
 at the matrix element level with the precuts
\mbox{$p_{T}^{l} > 200$~GeV},
 where $p_{T}^{l}$ denotes the transverse momentum for the three-momentum sum of the lepton momenta, 
 and $k_{T} > 140$ GeV for the jets.
We then simulate parton showering with a matching scale of 200~GeV,
 and study the detector effects.
Again, we have confirmed that the cross section after the final cut
 does not change much when the matching scale is varied  around 200~GeV by $\pm 60$~GeV.
The cross sections for $W(\rightarrow \tau \nu_{\tau})$ + jets events
 and $W(\rightarrow e \nu_{e} / \mu \nu_{\mu})$ + jets events 
 that pass the \textit{veryHighPT} cut are found to be
\begin{eqnarray}
\sigma_{ W(\rightarrow \tau \nu_{\tau})+{\rm jets} } \ (veryHighPT) &\simeq& 35.4 \ {\rm fb}\,,\\
\label{eq3}
\sigma_{ W(\rightarrow e \nu_{e} / \mu \nu_{\mu})+{\rm jets} } \ (veryHighPT) &\simeq& 22.2 \ {\rm fb}\,.
\label{eq4}
\end{eqnarray}

 In Table~1, we compare the numbers of background events estimated in Eqs.~(\ref{eq2}-\ref{eq4})
 with those estimated by the ATLAS collaboration~\cite{atlas} for an integrated luminosity of 1~fb$^{-1}$.
We find satisfactory agreement.
Following Ref.~\cite{fox}, we normalize our estimates
 for $Z$ + jets events and for $W$ + jets events
 to those of  the ATLAS collaboration.
The normalization factors are also shown in Table~1.
Note that the normalization factor for the  $Z(\nu \nu)$ + jets background (1.17)
 is consistent with Ref.~\cite{fox}, where it is estimated to be 1.2.

\begin{table}
\begin{center}
\begin{tabular}{|c|c|c|c|} \hline
Background & Our simulation & ATLAS collaboration & Normalization factor\\ \hline
$Z(\nu \nu)$ + jets & $106$ & $124 \pm 12 \pm 15$ & 1.17   \\ \hline
$W(\tau \nu)$ + jets & $35.4$ & $36 \pm 7 \pm 8$    & 1.08    \\ \cline{1-3}
$W(e/\mu \, \nu)$ + jets & $22.2$ & $26 \pm 4 \pm 3$ &   \\ \hline
\end{tabular}
\label{tab1}
\end{center}
\caption{
The numbers of background events that pass the \textit{veryHighPT} cut for $1$ fb$^{-1}$ LHC (at 7~TeV)
 according to our simulation (see Eqs.~\ref{eq2}-\ref{eq4}), compared to the estimates by ATLAS~\cite{atlas}.
The last column gives the ratios of the ATLAS estimate to our estimate.}
\end{table}

The DM signal in mono-jet events arises mainly from the following subprocesses:
\begin{eqnarray}
g \ u &\rightarrow& u \ \chi \ \bar{\chi} \ ,
\\
g \ d &\rightarrow& d \ \chi \ \bar{\chi} \ .
\end{eqnarray}
We simulate the DM contribution to mono-jet events by following the method we have adopted
  for the background simulations.
We generate $p p \rightarrow \chi \bar{\chi}$ + 1, 2 jets events
 at the matrix element level with the precuts \met $> 200$~GeV and $k_{T} > 140$~GeV for the jets.
Parton showering is simulated with a 200~GeV matching scale,
 and detector effects are simulated. 
We have also confirmed for the DM signal that the final results are not significantly affected by
 varying the matching scale around 200~GeV.
Finally, we rescale our signal by a factor of 1.17; see Table~1.

\begin{figure}[tbp]
  \begin{center}
   \includegraphics[width=100mm]{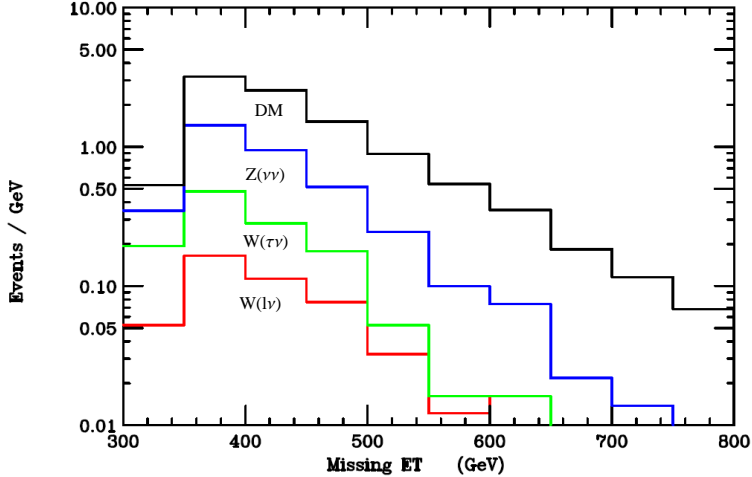}
  \end{center}
 \caption{The \met distribution of the backgrounds and the DM signal for 1~fb$^{-1}$
 at ATLAS. The curves from bottom to top successively include the 
 $W(e/\mu \, \nu)$ + jets background, the \mbox{$W(\tau \nu)$ + jets} background, the 
 $Z(\nu \nu)$ + jets background, and the DM signal with \mbox{$\Lambda=400$~GeV} and $m_{\chi}=10$~GeV.
{\it i.e.,} the curve labelled ``DM" corresponds to the sum of all the backgrounds and the DM signal.
  }
  \label{fig1}
\end{figure}

In order to compare our results with those of Ref.~\cite{fox},
 we estimate the DM signal with $\Lambda = 400$~GeV, $g_{U}=g_{D}=g_{Q}=1$ and $m_{\chi} = 10$~GeV.
We show the \met distribution in Fig.~\ref{fig1}.
All the curves are in good agreement with those in Fig. 2 of Ref.~\cite{fox}.
The difference in the \mbox{300-350~GeV} bin is likely
 due to the different modeling of parton showers.

To compare the lower bound on $\Lambda$ from our analysis with that from Ref.~\cite{fox},
 we derive the observed bound for $m_{\chi}=10$ GeV and $g_{U}=g_{D}=g_{Q}=1$.
It should be noted that only when all three couplings of the effective Lagrangian in Eq.~(\ref{lag}) are equal,
 do we have purely vector-like couplings for both up and down quarks, as assumed in Ref.~\cite{fox}.
Isospin violation with purely vector-like couplings is not compatible with electroweak symmetry.
To place 90\% C.L. lower limits on $\Lambda$ we require~\cite{fox}
\begin{eqnarray}
 \chi^{2} \ \equiv \ \frac{ (N_{obs} - N_{DM}(\Lambda) - N_{SM})^{2} }
{ N_{DM}(\Lambda) + N_{SM} + \sigma_{SM}^{2} }&<& 2.71  \ ,
\label{chi2}
\end{eqnarray}
 where $N_{obs}, N_{DM}$ and $N_{SM}$ are the number of observed events,
 the expected number of signal events, and the SM expectation, respectively, and
 $\sigma_{SM}$ is the uncertainty in $N_{SM}$.
We adopt $N_{obs}=167$ and $N_{SM}\pm \sigma_{SM}= 193\pm 25$ as obtained by the ATLAS collaboration in 1~fb$^{-1}$ of data.
 After rescaling, we find the number of signal events for $\Lambda=400$~GeV to be $N_{DM}=319$,
 which implies: $N_{DM}(\Lambda)=319 \times (400{\rm GeV} / \Lambda)^{4}$.
Finally, using the criterion of Eq.~(\ref{chi2}), we obtain the 90\% C.L. bound,
\begin{eqnarray}
\Lambda &>& 783 \ {\rm GeV} \ .
\label{bound}
\end{eqnarray}
On the other hand, Fig.~4 of Ref.~\cite{fox} shows separate bounds on $\Lambda$ for DM-up quark and
 DM-down quark interactions, which are both purely vector-like.
These bounds can be translated into a bound on $\Lambda$ for $g_{U}=g_{D}=g_{Q}=1$:
\begin{eqnarray}
\Lambda &>& (700^{4} + 575^{4})^{1/4} \ {\rm GeV} \ = \ 769 \ {\rm GeV} \ ,
\end{eqnarray}
 which is consistent with our estimate in Eq.~(\ref{bound}).
It is clear that our estimate of the number of signal events is consistent with that of Ref.~\cite{fox}.

For $g_{U}=1$ and $g_{D}=g_{Q}=0$, we derive the following bound on $\Lambda$
 using the data and the SM background estimates of the ATLAS collaboration~\cite{atlas}:
\begin{eqnarray}
\Lambda &>& 599 \ {\rm GeV} \ .
\end{eqnarray}

\subsection{Study for $\sqrt{s}=14$ TeV}

\ \ \ In this subsection, we study the DM contribution and the SM background to
 the mono-jet cross section at the LHC with $\sqrt{s}=14$ TeV.
We introduce the following cut, which we refer to as $MonoJ14TeV$:

\begin{itemize}
 \item Require \met $  >  800$ GeV.
 \item Highest $p_{T}$ jet should satisfy $\vert \eta_{1} \vert < 2.0$ and $p_{T 1} > 700$~GeV.
 \item Event is vetoed if the second hardest jet satisfies $\vert \eta_{2} \vert < 4.5$, and 
 $p_{T 2} > 120$~GeV or  
 $\Delta \phi (\vec{p}_{T 2}, \vec{E}_{{\rm T}}\hspace{-1.10em} \ \ \ ) < 0.5$.
 \item Event is vetoed if there is another jet that satisfies $\vert \eta \vert < 4.5$ and $p_{T} > 60$~GeV.
 \item Event is vetoed if there is an electron that satisfies $\vert \eta_{e} \vert < 2.47$ and $p_{T e} > 20$~GeV.
 \item Event is vetoed if there is a muon that satisfies $\vert \eta_{\mu} \vert < 2.4$ and $p_{T \mu} > 10$~GeV.
\end{itemize}
In comparison to the ATLAS \textit{veryHighPT} selection cut,
 we keep the lepton veto conditions and scale all the jet $p_{T}$ cuts by a factor of 2,
 while the \met cut is chosen to reduce the $Z(\nu \nu)$ + jet background 
 more efficiently in order to increase the signal significance
 for smaller DM signal cross sections.

We estimate the $Z(\nu \nu)$ + jet background
 in a manner similar to that for $\sqrt{s}=7$~TeV,
 by changing the precuts to \met$> 600$ GeV and $k_{T} > 240$ GeV for the jets,
 and the matching scale to $340$ GeV.
After testing the stability of our estimate under variation of the precuts and 
 the matching scale, we find that the $Z$ + jets background cross section is
\begin{eqnarray}
\sigma_{Z+jets} (MonoJ14TeV) &=& 12.8 \ {\rm fb} \ .
\label{zjets}
\end{eqnarray}
Similarly, we estimate the cross sections for the $W(\tau \nu)$ + jet and $W(l \nu)$ + jet
 backgrounds with the precuts $p_{T}^{l} > 600$ GeV and $k_{T} > 240$ GeV for the jets,
 and a matching scale of 340~GeV. We find
\begin{eqnarray}
\sigma_{W(\tau \nu)+jets} (MonoJ14TeV) &=& 2.4 \ {\rm fb} \ , \\ 
\sigma_{W(l \nu)+jets} (MonoJ14TeV) &=& 2.4 \ {\rm fb} \ .
\label{wjets}
\end{eqnarray}

To estimate the DM signal contribution,
 we set $(g_{Q}, g_{U}, g_{D})=(0, \cos \phi, \sin \phi)$
 so that the magnitudes of the couplings are given by $1/\Lambda^{2}$,
 and the isospin violation is parametrized by the phase $-\pi/2 < \phi \leq \pi/2$.
The up and down quark couplings have the same sign for $0 < \phi <\pi/2$,
 and opposite signs for $-\pi/2 < \phi < 0$.
We repeat the simulation steps of the $\sqrt{s}=7$~TeV case,
 by changing the precuts to \met $> 600$ GeV and $k_{T} > 240$~GeV for the jets,
 and the matching scale to 340~GeV.
We calculate the cross sections for various values of the DM mass $m_{\chi}$ and $\phi = \tan^{-1}(g_D/g_U)$. The results 
 for three representative couplings, $\phi=0 \ (g_U=1)$, $\phi=\pm \pi/4 \ (\vert g_U \vert = \vert g_D \vert = 1/\sqrt{2})$
 and \mbox{$\phi=\pi/2 \ (g_D=1)$} are listed in Table~2 in units of $(800\, {\rm GeV} / \Lambda)^{4}$~fb.
The mono-jet cross section does not distinguish between the signs of the $g_U$ and $g_D$ couplings,
 and hence $\pm \phi$ give the same prediction.
\begin{table}
\begin{center}
\begin{tabular}{|c|c|c|c|} \hline
$m_{\chi}$ & $\phi=0$ & $\phi=\pm \pi/4$ & $\phi=\pm \pi/2$ \\ \hline
10 GeV  & 15.9 & 11.0 & 6.37    \\ \hline
300 GeV & 15.1 & 10.5 & 5.91    \\ \hline
500 GeV & 13.0 & 8.89 & 4.98    \\ \hline
\end{tabular}
\end{center}
\caption{
DM signal cross sections (in units of $(800\, {\rm GeV} / \Lambda)^{4}$~fb) for $\sqrt{s}=14$ TeV with the $MonoJ14TeV$ cut,
 for various values of the DM mass $m_{\chi}$ and the ratio of the DM couplings parametrized by
 $\phi=\tan ^{-1}(g_{D}/g_{U})$.}
\end{table}
\\

\section{Mono-photon + \met}

\ \ \ The mono-jet cross section alone does not contain information on $\phi$.
Therefore additional signals are needed, and we study the mono-photon plus \met signal
 in this section.
Since the photon coupling to the up quark is twice that to the down quark,
 we expect a strong dependence of the DM signal on $\vert \phi \vert$.
We first focus on the $\sqrt{s}=7$~TeV case
 and reproduce the results of Ref.~\cite{fox}.
We then proceed to the $\sqrt{s}=14$~TeV case, 
 and look for an appropriate selection cut at high energies and at high integrated luminosity.

\subsection{Study for $\sqrt{s}=7$ TeV}

\ \ \ Following Refs.~\cite{fox, cms}, we implement the following selection cut,
 which we call $MonoG7TeV$:

\begin{itemize}
 \item Require a photon with $\vert \eta_{\gamma} \vert < 1.44$ and $p_{T \gamma} > 95$~GeV.
 \item Require \met $ > 80$~GeV.
 \item Event is vetoed if there is a jet with $\vert \eta_{j} \vert < 3.0$ and $p_{T j} > 20$~GeV.
 \item Event is vetoed if there is an isolated lepton with $p_{T l} > 10$ GeV and
 $\Delta R(l,\gamma) \equiv \sqrt{ (\eta_l-\eta_\gamma)^{2} + (\phi_l-\phi_\gamma)^{2} } > 0.04$.
\end{itemize}

The leading background arises from $Z(\nu \nu)$ + $\gamma$ events.
In addition, $Z(\nu \nu)$ + jet events with the jet misidentified as a photon,
 and $W(e \nu)$ events with the electron misidentified as a photon
 also contribute to the background.

We simulate the $Z(\nu \nu)$ + $\gamma$ background as follows.
We first generate $Z(\nu \nu)$ + $\gamma$ and $Z(\nu \nu)$ + $\gamma$ + 1 jet
 events at the matrix element level with the precuts 
 $p_{T\gamma} > 60$~GeV for the photon and $k_{T} > 60$~GeV for the jet,
 and simulate parton showering by using $k_{T}$-jet matching
 with a matching scale of $84$ GeV.
After simulating the detector effects with PGS and imposing the selection cut,
 we obtain
\begin{eqnarray}
\sigma_{Z\gamma}( MonoG7TeV ) &=& 48.0 \ {\rm fb} \ .
\end{eqnarray}
This gives 54.7 events in 1.14~fb$^{-1}$ of data,
 while the CMS collaboration estimates 36.4 events~\cite{cms}.
We therefore multiply the DM signal (which has a similar event topology as the $Z(\nu \nu)$ + $\gamma$ background) by the ratio $36.4/54.7 = 0.67$.
This rescaling reflects the additional criteria for photon indentification
 specific to the CMS detector.
Note that the factor of 0.67 is in agreement with the analysis in Ref.~\cite{fox},
 which finds 0.71.

The DM signal arises mainly from the following subprocesses:
\begin{eqnarray}
u \ \bar{u} &\rightarrow& \gamma \ \chi \ \bar{\chi} \ ,
\\
d \ \bar{d} &\rightarrow& \gamma \ \chi \ \bar{\chi} \ .
\end{eqnarray}
Setting $m_{\chi}=10$ GeV and $g_{U}=g_{D}=g_{Q}=1$,
 we generate $\chi \bar{\chi}$ + $\gamma$ and $\chi \bar{\chi}$ + $\gamma$ + 1 jet
 events at the matrix element level with the same precuts as for the background,
 which are then processed to the parton showering simulation with the same matching scale
 and finally the detector simulation is done using PGS.
In this way, we find the DM signal cross section before the rescaling to be
\begin{eqnarray}
\sigma_{DM \gamma}( MonoG7TeV ) &=& 63.3 \times ( 400\ {\rm GeV} / \Lambda )^{4} \ {\rm fb} \ .
\label{DMG}
\end{eqnarray}

To compare our analysis with that of Ref.~\cite{fox}, we estimate the 90\% C.L. lower bound on $\Lambda$
 by using the criterion of Eq.~(\ref{chi2}).
For $N_{SM}$ and $\sigma_{SM}$, we adopt the numbers estimated by the CMS collaboration,
 which are $N_{SM}=67.3$ and $\sigma_{SM}=8.4$ in 1.14~fb$^{-1}$~\cite{cms}.
The CMS collaboration reports $N_{obs}=80$~\cite{cms}.
$N_{DM}$ is estimated by multiplying Eq.~(\ref{DMG}) by the normalization factor, 0.67.
We obtain the 90\% C.L. bound,
\begin{eqnarray}
\Lambda &>& 422 \ {\rm GeV} \ .
\label{L}
\end{eqnarray}
On the other hand, Fig.~8 of Ref.~\cite{fox} shows separate bounds on DM-up quark
 and DM-down quark interactions.
These bounds can be translated into a bound on $\Lambda$ for the case, $g_{U}=g_{D}=g_{Q}=1$:
\begin{eqnarray}
\Lambda &>& (400^{4} + 240^{4})^{1/4} \ {\rm GeV} \ = \ 412 \ {\rm GeV} \ .
\end{eqnarray}
This is consistent with our result in Eq.~(\ref{L}).

\subsection{Study for $\sqrt{s}=14$ TeV}

\ \ \ We introduce the following selection cut for the LHC with $\sqrt{s}=14$ TeV,
 which we name $MonoG14TeV$-$a$:
\begin{itemize}
 \item Require a photon with $\vert \eta_{\gamma} \vert < 1.44$ and $p_{T \gamma} > 140$~GeV.
 \item Require \met $ > 140$~GeV.
 \item Event is vetoed if there is a jet with $\vert \eta_{j} \vert < 3.0$ and $p_{T j} > 40$~GeV.
 \item Event is vetoed if there is an isolated lepton with $p_{T l} > 10$~GeV and $\Delta R(l,\gamma) > 0.04$.
\end{itemize}
We also introduce the selection cut $MonoG14TeV$-$b$ which requires 
 ($p_{T \gamma}$, $E_{{\rm T}}$\hspace{-1.10em}/ \ ) $ > (200, 200)$~GeV,
 and the cut $MonoG14TeV$-$c$ which requires ($p_{T \gamma}$, $E_{{\rm T}}$\hspace{-1.10em}/ \ ) $ > (260, 260)$~GeV,
 and for which all the other requirements are the same as for $MonoG14TeV$-$a$.

We estimate the $Z(\nu \nu)$ + $\gamma$ background in a similar manner
 to the case of $\sqrt{s}=7$~TeV, by changing the precuts to 
 $p_{T\gamma} > 100$~GeV for the photon and $k_{T} > 100$~GeV for the jet,
 and the matching scale to 140~GeV.
We estimate the DM signal contribution
 with the same precuts and matching scale as for
 the $Z(\nu \nu)$ + $\gamma$ background estimation for $\sqrt{s}=14$ TeV.

Table 3 shows the cross sections and significance factors ($S/\sqrt{S+B}$)
 for the DM signal with $m_{\chi}=10$~GeV, $\Lambda=800$~GeV and $\phi=\pi/4$, 
 where the three selection cuts are imposed.
The numbers are in units of fb for the cross sections and fb$^{1/2}$ for the significance factor.
\begin{table}
\begin{center}
\begin{tabular}{|c|c|c|c|} \hline
Cut                  & $\sigma_{Z \gamma}$  & $\sigma_{DM \gamma}$ & $\sigma_{DM \gamma}/\sqrt{\sigma_{Z \gamma}+\sigma_{DM \gamma}}$ \\ \hline
$MonoG14TeV$-$a$     & 35.8                 & 5.49                 & 0.85    \\ \hline
$MonoG14TeV$-$b$     & 11.6                 & 3.34                 & 0.87    \\ \hline
$MonoG14TeV$-$c$     & 4.51                 & 2.13                 & 0.83    \\ \hline
\end{tabular}
\end{center}
\caption{ Cross sections (in fb) and significance factors (fb$^{1/2}$) 
 for the DM signal with \mbox{$m_{\chi}=10$~GeV}, $\Lambda=800$~GeV and $\phi=\pi/4$.
 The three selection cuts for mono-photon events are compared.}
\end{table}
We find that the significance factor does not depend much on the choice of the selection cut.
This is because the background cross section decreases by a factor of 8
 and the signal also decreases by $2.6 \simeq \sqrt{8}$
 when we change the cut from $MonoG14TeV$-$a$ to $MonoG14TeV$-$c$.
For our study, we adopt the selection cut $MonoG14TeV$-$a$ which yields the largest signal cross section.
This is because, with more events, it is easier to discriminate the signal from the background
 from the difference in their \met and $p_{T\gamma}$ distributions
 as well as from the correlation between $\gamma$ and \met momenta,
 although we do not perform such an analysis in this paper.

\begin{figure}[tbp]
  \begin{center}
   \includegraphics[width=85mm]{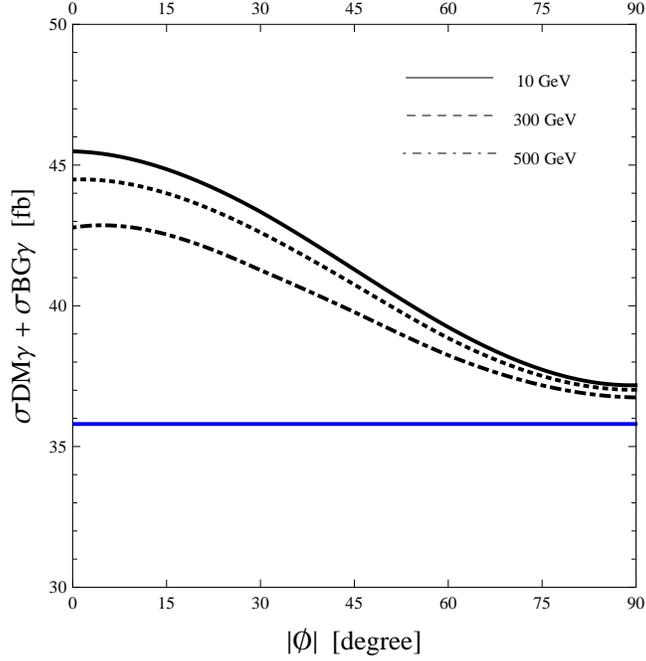}
  \end{center}
 \caption{The $Z$ + $\gamma$ background cross section and the DM signal cross section
  for $\Lambda=800$ GeV that satisfy the $MonoG14TeV$-$a$ criteria,
  as a function of $\vert \phi \vert$ for  the couplings
  $(g_Q, g_U, g_D)=(0, \cos \phi, \sin \phi)$.
 The horizontal solid line corresponds to the background cross section.
 The curves correspond to the
  background + signal cross sections for three values of $m_{\chi}$.
  }
  \label{fig2}
\end{figure}

In Fig.~\ref{fig2}, we show the cross sections for the $Z$ + $\gamma$ background and 
 the DM signal for $\Lambda=800$ GeV and $(g_{Q}, g_{U}, g_{D})=(0, \cos \phi, \sin \phi)$
 that satisfy the cut $MonoG14TeV$-$a$.
The three curves correspond to $m_{\chi}=10, \ 300, \ 500$~GeV.
The mono-photon cross section does not depend on the signs of the $g_U$ and $g_D$ couplings,
 and hence $\pm \phi$ give the same prediction.

In Table~4, we present the DM signal cross sections in units of $(800\, {\rm GeV}/\Lambda)^{4}$~fb
 for $\phi=0, \ \pm \pi/4$  
 and \mbox{$\pi/2$}.
\begin{table}
\begin{center}
\begin{tabular}{|c|c|c|c|} \hline
$m_{\chi}$ & $\phi=0$ & $\phi=\pm \pi/4$ & $\phi=\pm \pi/2$ \\ \hline
10 GeV  & 9.69 & 5.49 & 1.38    \\ \hline
300 GeV & 8.69 & 4.96 & 1.22    \\ \hline
500 GeV & 6.98 & 3.98 & 0.945    \\ \hline
\end{tabular}
\end{center}
\caption{DM signal cross sections (in units of $(800\, {\rm GeV} / \Lambda)^{4}$~fb) for $\sqrt{s}=14$ TeV that satisfy the $MonoG14TeV$-$a$ cut.
}
\end{table}
Roughly speaking, the $\phi=\pi/2 \ (g_D=1)$ case gives about a factor of 7 smaller DM signal cross section than
 the $\phi=0 \ (g_U=1)$ case, 
 while the cross section in the $\phi=\pm \pi/4 \ (\vert g_U \vert = \vert g_D \vert = 1/\sqrt{2})$ case
 is about half of that in the $\phi=0$ case,
 for all $m_{\chi}$ between 10~GeV and 500~GeV.
This reflects the combined effect of the QED coupling ratio $(Q_d/Q_u)^{2}=1/4$ and
 the ratio of the down quark to up quark parton distribution functions in $pp$ collisions.

\begin{figure}[tbp]
  \begin{center}
   \includegraphics[width=85mm]{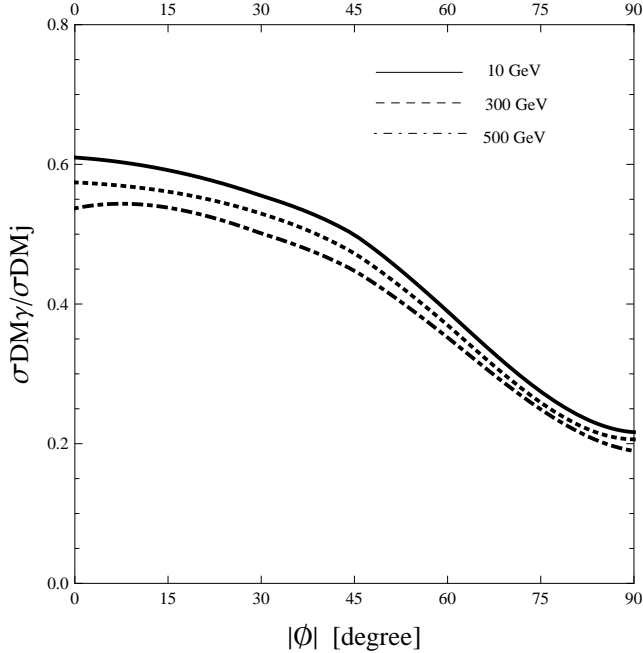}
  \end{center}
 \caption{Ratio of the DM signal cross section in the mono-photon channel with the $MonoG14TeV$-$a$ cut
 to the cross section in the mono-jet channel with the $MonoJ14TeV$ cut 
 for three values of $m_\chi$.
  }
  \label{fig3}
\end{figure}

In Fig.~\ref{fig3}, we show the ratio of the DM signal cross sections
 in the mono-photon channel with the $MonoG14TeV$-$a$ cut
  to the cross section in the mono-jet channel with the $MonoJ14TeV$ cut.
We find that this ratio is very sensitive to 
 the coupling ratio $\vert g_D/g_U \vert = \tan \vert \phi \vert$,
 but insensitive to the DM mass $m_{\chi}$.
It is independent of $\Lambda$ and $\sqrt{g_{U}^{2}+g_{D}^{2}}$.

Experimentally, the cross section ratio, $\sigma_{DM\gamma}/\sigma_{DMj}$,
 is determined by
\begin{eqnarray}
\frac{ \sigma_{DM\gamma}({\rm monophoton \ cut}) }
{ \sigma_{DMj}({\rm monojet \ cut}) }
&=& \frac{ (N^{\gamma}_{obs} - N^{\gamma}_{SM})({\rm monophoton \ cut})  }
{ (N^{j}_{obs} - N^{j}_{SM})({\rm monojet \ cut}) } \ ,
\end{eqnarray}
 where $N^{\gamma}_{obs}$ and $N^{j}_{obs}$  are the observed number of mono-photon and mono-jet events, respectively, 
 and $N^{\gamma}_{SM}$ and $N^{j}_{SM}$ are  the SM expectations for
 mono-photon background events and  mono-jet background events, respectively.
Since the cross section ratio has a common value for a wide range of $m_{\chi}$,
 we can determine the value of $\vert \phi \vert$
 by comparing the observed value of the cross section ratio with Fig.~\ref{fig3}.

As an illustration, we estimate the integrated luminosity $L$ needed to measure the cross section ratio
 with 10\% accuracy close to the point $\vert \phi \vert = \pi/4$
 for $\Lambda=800$~GeV and $m_{\chi}=10$~GeV.
We assume that the statistical uncertainty of the number of events, $N$,
 follows $\Delta N = \sqrt{N}$,
 and ignore systematic uncertainties.
To estimate $N^{\gamma}_{obs}$ and $N^{\gamma}_{SM}$,
 we rescale the cross sections by the normalization factor 0.67 obtained in Section 4.1.
The statistical uncertainty of $\sigma_{DM \gamma}/\sigma_{DMj}$ is given by
\begin{eqnarray}
\Delta \left( \frac{\sigma_{DM\gamma}}{\sigma_{DMj}} \right) &=& \left( \frac{\sigma_{DM\gamma}}{\sigma_{DMj}} \right)
 \ \sqrt{ \frac{ N^{j}_{obs}+N^{j}_{SM} }{ (N^{j}_{obs} - N^{j}_{SM})^{2} }
\ + \ \frac{ N^{\gamma}_{obs}+N^{\gamma}_{SM} }{ (N^{\gamma}_{obs} - N^{\gamma}_{SM})^{2} }
} \ .
\label{uncert}
\end{eqnarray}
Inserting $N_{SM}^{\gamma} = L \times 0.67 \times 35.8$~fb and
 $N_{obs}^{\gamma}-N_{SM}^{\gamma} = L \times 0.67 \times 5.49$~fb from Table~3,
 $N_{SM}^{j} = L \times (12.8+2.4+2.4)$ fb~from Eqs.~(\ref{zjets}-\ref{wjets}),
 and $N_{obs}^{j}-N_{SM}^{j} = L \times 11.0$ fb from Table~2,
we find that in order to obtain a 10\% measurement of the monophoton-to-monojet cross section ratio,
$L \ \simeq \ 420 \ {\rm fb}^{-1}$ is needed.

\begin{figure}[tbp]
  \begin{center}
   \includegraphics[width=120mm]{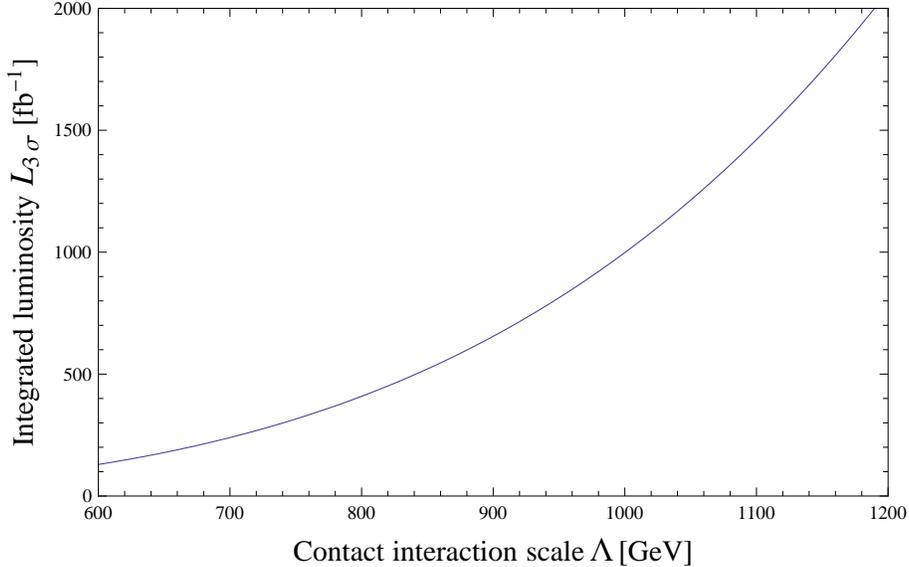}
  \end{center}
 \caption{
 Integrated luminosity $L_{3\sigma}$ needed to reject the isospin-conservation hypothesis,
  $\phi=\pi/4 \, (g_U=g_D=1/\sqrt{2})$, at 3$\sigma$
  given that $\phi=0 \, (g_U=1, \, g_D=0)$. Here, $m_\chi=10$~GeV.
 Only statistical uncertainties are taken into account.
 }
 \label{luminosityneeded}
\end{figure}

We now evaluate the integrated luminosity needed to establish the isospin violating nature of
 the DM-quark couplings as a function of $\Lambda$.
As an example, consider the ability to reject the hypothesis that isospin is conserved \textit{i.e.},
 $\phi = \pi/4 \, (g_U=g_D=1/\sqrt{2})$, given that
 $\phi = 0 \, (g_U=1, \, g_D=0)$. 
The DM mass $m_{\chi}$ is fixed at 10~GeV.
To reject the hypothesis at 3$\sigma$, from Fig.~\ref{fig3}
 we require
\begin{align}
\Delta \left( \frac{\sigma_{DM\gamma}}{\sigma_{DMj}} \right)(\phi=0) \ &= \ 
\frac{1}{3} \left(
\ \frac{\sigma_{DM\gamma}}{\sigma_{DMj}}(\phi=0) - \frac{\sigma_{DM\gamma}}{\sigma_{DMj}}(\phi=\pi/4) \
\right)
\nonumber \\
&= \ 0.037\,. \label{uncert required}
\end{align}
Then using Eqs.~(\ref{uncert}) and~(\ref{uncert required}),
in Fig.~\ref{luminosityneeded}
we show the integrated luminosity needed to reject the isospin-conservation hypothesis at the 3$\sigma$ C.L.,
 $L_{3\sigma}$, as a function of the contact interaction scale $\Lambda$.

\begin{figure}[tbp]
  \begin{center}
   \includegraphics[width=120mm]{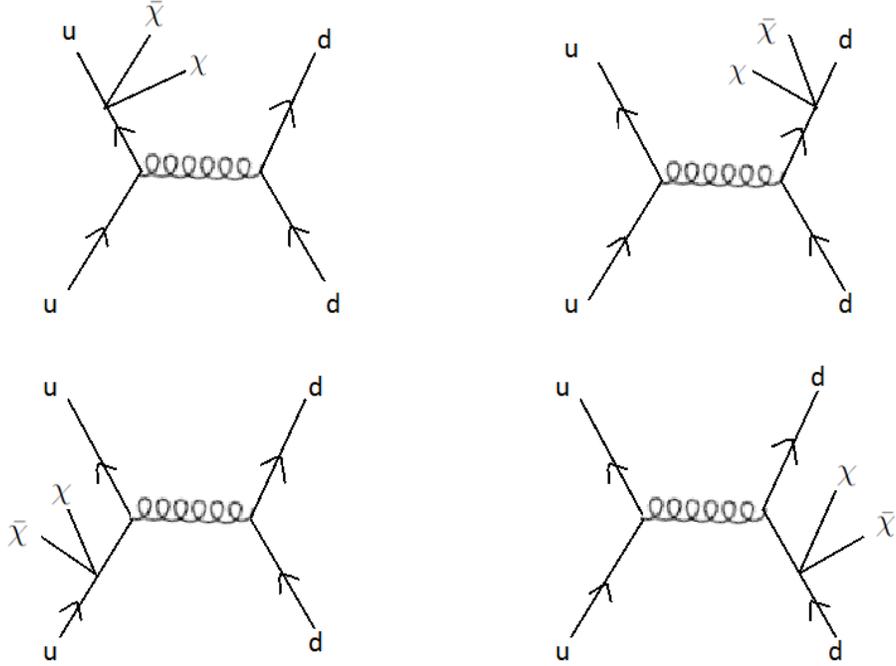}
  \end{center}
 \caption{Feynman diagrams for the subprocess, 
  $u_R d_R \rightarrow u_R d_R \chi \bar{\chi}$.
 The diagrams on the left are proportional to $g_U$,
  and the ones on the right are proportional to $g_D$.
 }
 \label{fig4}
\end{figure}

\section{Di-jet + \met}

\ \ \ Since the mono-jet and mono-photon cross sections depend only on the absolute values
 of $g_{U}$ and $g_{D}$,
 a measurement of the relative sign of $g_{U}$ and $g_{D}$ is not possible from these channels.
To determine the relative sign, we now focus on events with two hard jets and large \mbox{\met.}
This channel can be sensitive to the relative sign of the
 $g_{U}$ and $g_{D}$ couplings
 because in the subprocess,
\begin{eqnarray}
u_R d_R &\rightarrow& u_R d_R \chi \bar{\chi} \ ,
\label{dijet}
\end{eqnarray}
 the amplitudes where $\chi \bar{\chi}$ are emitted from the up quark interfere with
 the amplitudes where $\chi \bar{\chi}$ are emitted from the down quark,
 as can be seen from the Feynman diagrams in Fig.~\ref{fig4}.
The interference term is directly proportional to $g_{U}g_{D}$ so that
 the cross section depends on the sign of $g_{U}g_{D}$.

%

\subsection{Extracting the Interference}

\ \ \  In this subsection we discuss kinematic conditions that enhance 
interference effects in the diagrams of Fig.~\ref{fig4}.


\subsubsection{Suppressing contributions from quark-gluon collisions}

First, it is necessary to suppress large contributions from 
  quark-gluon collisions,
\begin{eqnarray}
q \ g &\rightarrow& q \ g \ Z(\nu \nu) \ , \label{sub1}
\\
q \ g &\rightarrow& q \ g \ \chi \bar{\chi} \ , \ \ \ (q=u, \ d). \label{sub2}
\end{eqnarray}
While the subprocesses in Eq.~(\ref{sub1}) give the dominant background for the DM plus di-jet
 production events, the subprocesses in Eq.~(\ref{sub2}) should also be suppressed
 because their cross sections do not depend on $g_U g_D$.
The contributions from quark-gluon collisions can be reduced,
 compared to the signal subprocess in Eq.~(\ref{dijet}) from quark-quark collisions
 by imposing a very large cut on the ``di-jet cluster transverse mass"~\cite{cluster MT}:
\begin{eqnarray}
M_T(jj; E_{{\rm T}}\hspace{-1.10em}/ \ \ ) &\equiv& \sqrt{ M_{12}^{2} + \vert \vec{p}_{1T} + \vec{p}_{2T} \vert^2 } \ + \ E_{{\rm T}}\hspace{-1.10em}/ \ \ \ ,
\end{eqnarray}
 where $M_{12}$ is the invariant mass of the two hardest jets,
 and $\vec{p}_{1T}+\vec{p}_{2T}$ is 
 the vector sum of the transverse momenta of the two hardest jets.
Since the parton center-of-mass energy ($\sqrt{\hat{s}}$) is always larger than 
 $M_T(jj; E_{{\rm T}}\hspace{-1.10em}/ \ \ )$,
 by imposing a cut on $M_T(jj; E_{{\rm T}}\hspace{-1.10em}/ \ \ )$,
 we can select those events that satisfy
\begin{eqnarray}
\sqrt{\hat{s}} & \geq & M_T(jj; E_{{\rm T}}\hspace{-1.10em}/ \ \ ) \ > \ 
 (M_T(jj; E_{{\rm T}}\hspace{-1.10em}/ \ \ ))^{cut} \ .
\end{eqnarray}
By choosing a large value for $(M_T(jj; E_{{\rm T}}\hspace{-1.10em}/ \ \ ))^{cut}$,
 we can suppress the contributions from quark-gluon collisions
 relative to those from quark-quark collisions,
 because the former subprocesses have relatively less parton center-of-mass energy.
We note that the contributions from the same quark collisions:
\begin{eqnarray}
q \ q &\rightarrow& q \ q \ Z(\nu \nu) \ ,
\\
q \ q &\rightarrow& q \ q \ \chi \bar{\chi} \ , \ \ \ (q=u, \ d), 
\end{eqnarray}
 cannot be suppressed by a cut on $M_T(jj; E_{{\rm T}}\hspace{-1.10em}/ \ \ )$.
We therefore treat them as irreducible backgrounds for 
 a measurement of the sign of $g_U g_D$.

\subsubsection{Utilizing the generalized null radiation zone theorem}

\ \ \ 
Generally speaking,
 when a subprocess contains two Feynman diagrams that interefere,
 the ratio of the interference term to the cross section is maximized
 when the amplitudes of the two diagrams take the same absolute value.
The null radiation zone theorem~\cite{null zone} provides us with 
 powerful criteria for identifying such kinematic regions.
The theorem states that
 for any tree-level Feynman diagram,
 if the ratio $Q_{k}/(p_{k} \cdot q)$ (where $Q_{k}$ is the charge of an external particle, 
 $p_{k}$ is its four-momentum, and $q$ is the four-momentum of the emitted photon)
 is the same for all external particles
 and if the charge $Q_{k}$'s are conserved,
then the sum of the amplitudes of all the tree-level diagrams 
 made by adding one photon emission vertex to the original diagram
 vanishes for all helicities.
Since the amplitude for each diagram does not vanish,
 the theorem implies that the amplitudes of the contributing diagrams
 cancel exactly, {\it i.e.}, they interfere maximally.
The original theorem applies to massless vector boson emission 
 via a vector coupling to fermions.
We make use of the theorem for our dark matter pair which couples to quarks
 via vector and axial vector couplings.
Since the dark matter pair is not massless,
 we first generalize the theorem to  emission of a massive vector current.

We now show that the null radiation zone theorem can be generalized to emission of 
 a massive vector boson,
 or a vector current whose invariant mass squared is time-like,
 $q^2 > 0$.
Consider the following process in which a neutral vector current, $V$, is emitted:
\begin{eqnarray}
a \ + \ b &\rightarrow& 1 \ + \ 2 \ + \ ... \ + n \ + \ V \ .
\end{eqnarray}
We label the particle four-momenta and charges in the initial state
 by $p_i, \ Q_i$ $(i=a,b)$, 
 and those in the final state by $p_f, \ Q_f$ $(f=1,2,...,n)$.
We denote the four-momentum of the vector current by $q$.
The tree-level scattering amplitude should vanish for all helicities
 when the following conditions are satisfied:
\begin{eqnarray}
\frac{Q_i}{2 p_i \cdot q - q^2} &=& \frac{Q_f}{2 p_f \cdot q + q^2} \ = \ {\rm (a \ common \ value)}
\ \ \ {\rm for \ all} \ i \ {\rm and} \ f \ ,
\\
\sum_i Q_i &=& \sum_f Q_f \ .
\end{eqnarray}

Let us focus on the four diagrams of Fig.~\ref{fig4}.
We denote the four-momenta of the incoming quarks by $k_1$ and $k_2$,
 those of the outgoing quarks by $p_1$ and $p_2$
 and the four-momentum sum of the DM momenta by $q$.
We notice that the values of $2 p_1 \cdot q + q^2$ and $2 p_2 \cdot q + q^2$
 are always positive.
The values of $2 k_1 \cdot q - q^2$ and $2 k_2 \cdot q - q^2$ can take both signs,
 but their sum is always positive because $k_1+k_2=p_1+p_2+q$.
Therefore the null radiation zone can be realized only when $g_U$ and $g_D$
 take the same sign.
We thus expect that the interference in $pp$ collisions
 is destructive if $g_U$ and $g_D$ take the same sign,
 and is constructive otherwise.

The theorem suggests that if the condition,
\begin{eqnarray}
k_1 \cdot q - \frac{q^2}{2} &=& k_2 \cdot q - \frac{q^2}{2} 
\ = \ p_1 \cdot q + \frac{q^2}{2} \ = \ p_2 \cdot q + \frac{q^2}{2}\,,
\label{nrzt}
\end{eqnarray}
 is satisfied,
 the amplitudes of the four diagrams of Fig.~\ref{fig4} cancel completely for $g_U=g_D$,
 while the sum of the amplitudes is maximally enhanced for $g_U=-g_D$.
Equation~(\ref{nrzt}) reduces to the following kinematic conditions
 for massless partons satisfying $p_1^2=p_2^2=k_1^2=k_2^2=0$:
\begin{eqnarray}
\vert \vec{p}_1 \vert &=& \vert \vec{p}_2 \vert \ ,  \label{massless}\\
q^2 &=& 0 \ ,
\end{eqnarray}
 where $\vert \vec{p}_1 \vert$ and $\vert \vec{p}_2 \vert$ 
 are respectively the magnitudes of the three-momenta of
 $p_1$ and $p_2$ in the \textit{colliding parton center-of-mass} frame.
Since our DM particle is not massless,
 the condition Eq.~(\ref{nrzt}) cannot be satisfied
 and the null radiation zone does not exist in the physical region.
However, we expect that strong destructive interference occurs
 for $g_U=g_D$ when the condition Eq.~(\ref{nrzt}) is approximately satisfied.

At hadron colliders, we cannot measure the missing mass $\sqrt{q^2}$,
 and cannot determine the colliding parton center-of-mass frame.
Therefore, to
 enhance kinematic regions around the null radiation zone,
 we must make use of the constraints among jet transverse momenta
 coming from Eq.~(\ref{massless}).
In the following we examine the selection cut,
\begin{eqnarray}
\vert p_{T1} - p_{T2} \vert &<& C \ p_{T1} \ ,
\end{eqnarray}
 where $p_{T1}$ and $p_{T2}$ are the transverse momenta
 of the hardest and second hardest jets, respectively, and $C$ is a number less than unity.
Although no selection cut can be applied to enhance kinematic regions
 that nearly satisfy Eq.~(\ref{nrzt}),
 such regions are automatically favored in $pp$ collisions with a fixed cut
 because they are the regions where the parton center-of-mass energy is minimized.

\subsection{Analysis}

\ \ \ The ``signal" in this analysis is not DM production itself,
 but the difference in the DM signal cross sections for $g_{U}/g_{D} > 0$
 and $g_{U}/g_{D} < 0$.
In this analysis, therefore, we consider the value of $\sigma_{DMjj}(\phi=-\pi/4)-\sigma_{DMjj}(\phi=\pi/4)$
 as the signal and treat the value of $\sigma_{DMjj}(\phi=-\pi/4)+\sigma_{SM}$ as the background.

We first examine selection cuts dubbed $DiJ$-$a$, $DiJ$-$b$, $DiJ$-$c$ and $DiJ$-$d$.
The $DiJ$-$a$ cut is:
\begin{itemize}
 \item Require two jets with $\vert \eta \vert < 4.5$ and $p_{T} > 200$~GeV.
 \item Require \met $ > 300$~GeV.
 \item Require
 %
 $M_T(jj; E_{{\rm T}}\hspace{-1.10em}/ \ \ ) = \sqrt{ M_{12}^{2} + \vert \vec{p}_{1T} + \vec{p}_{2T} \vert^2 } \ + \ E_{{\rm T}}\hspace{-1.10em}/ \  
 \ > \ 2 \ {\rm TeV}.$
 %
 \item Event is vetoed if there is another jet that satisfies $\vert \eta \vert < 4.5$ and $p_{T} > 100$~GeV.
 \item Event is vetoed if there is a jet whose three-momentum $\vec{p}$ satisfies
  $\Delta \phi (\vec{p}, \vec{E}_{{\rm T}}\hspace{-1.10em}/ \ \ ) < 0.2$.
\end{itemize}
The first, second and fourth conditions define the two jets plus large \met events that we study.
It is the third condition that enhances events from quark-quark collisions
 over those from quark-gluon collisions.
The last condition is necessary to reduce those $Z(\nu \nu)$ + jets events
 in which a quark emits the $Z$ boson almost collinearly;
 the amplitude receives collinear enhancement even for the $Z$ boson
 because $E_{jet} / M_Z \gg 1$.
For comparison, we consider a selection cut $DiJ$-$b$ that requires 
 $M_T(jj; E_{{\rm T}}\hspace{-1.10em}/ \ \ )> 2.5$ TeV, 
 and a selection cut $DiJ$-$c$ that requires
 $M_T(jj; E_{{\rm T}}\hspace{-1.10em}/ \ \ ) > 3$ TeV, 
 with the other conditions the same as for $DiJ$-$a$.
Based on the null radiation zone theorem,
 we also consider a selection cut $DiJ$-$d$ that requires
 \begin{eqnarray}
\vert p_{T1} - p_{T2} \vert &<& 0.5 \ p_{T1}\,,
\label{dijd}
\end{eqnarray}
 in addition to the conditions of $DiJ$-$b$. 
 ($p_{T1}$ and $p_{T2}$ are the transverse momenta of the hardest
 and second hardest jets, respectively.)
The cuts are summarized as follows:
\begin{eqnarray}
DiJ-a &:& M_T(jj; E_{{\rm T}}\hspace{-1.10em}/ \ \ ) \ > \ 2 \ {\rm TeV} \ . \\
DiJ-b &:& M_T(jj; E_{{\rm T}}\hspace{-1.10em}/ \ \ ) \ > \ 2.5 \ {\rm TeV} \ . \\
DiJ-c &:& M_T(jj; E_{{\rm T}}\hspace{-1.10em}/ \ \ ) \ > \ 3 \ {\rm TeV} \ . \\
DiJ-d &:& M_T(jj; E_{{\rm T}}\hspace{-1.10em}/ \ \ ) \ > \ 2.5 \ {\rm TeV} 
\ \ \ {\rm and} \ \ \ \vert p_{T1} - p_{T2} \vert \ < \ 0.5 \ p_{T1} \ .
\end{eqnarray}

Our simulation of the SM background only accounts for the dominant background from $Z(\nu \nu)$ + jets events.
To estimate the background cross section,
 we generate $Z (\nu \nu)$ + 2, 3 jets events at the matrix element level
 with the following precuts:
$E_{{\rm T}}\hspace{-1.10em}/ \ \ > \ 200$~GeV, 
$k_T \ > \ 160$~GeV,
$\Delta R( \vec{p}_j, \vec{p}_Z )  > 0.2$,
$\sqrt{\hat{s}}  >  2 \ (2.5)$ TeV when using the $DiJ$-$a$ cut 
($DiJ$-$b$, $DiJ$-$c$ and $DiJ$-$d$ cuts),
 where $\vec{p}_j$ is the three-momentum of any jet and
 $\vec{p}_Z$ that of the $Z$ boson.
We match the matrix element events with the parton shower
 with a matching scale of $220$~GeV,
 and perform a detector simulation.
We find that the cross section after the final cut
 does not change drastically when the matching scale is varied by $\pm 40$~GeV.

The DM production cross section is estimated in a similar manner;
 we generate $\chi \bar{\chi}$ + 2, 3 jets events
 with the same precuts as the background simulation
 ($\vec{p}_Z$ is replaced with the three-momentum sum of the DM momenta),
 and process them to parton showering and detector simulation
 with the same matching scale.

Table 5 shows the cross sections and the significance factors ($S/\sqrt{S+B}$)
 for  DM production with $m_{\chi}=10$~GeV, $\Lambda=800$~GeV and $\phi=\pm \pi/4$, 
 for the four selection cuts.
Also shown is the ratio of the interference effect 
 ($\sigma_{DM jj}(\phi=-\pi/4) - \sigma_{DM jj}(\phi=+\pi/4)$)
 to the DM production cross section for $\phi=-\pi/4$ ($\sigma_{DM jj}(\phi=-\pi/4)$).
Notice that in this analysis the ``signal" cross section
 corresponds to the difference between the DM cross sections
 for $\phi=\pi/4$ and $\phi=-\pi/4$.
The numbers are in units of fb for the cross sections 
 and fb$^{1/2}$ for the significance factors.
From Table 5
 we see that the $DiJ$-$b$ cut gives the largest significance factor.
We note in passing that the requirement Eq.~(\ref{dijd}) of the $DiJ$-$d$ cut,
 which is based on the null radiation zone theorem,
 does enhance the ratio of the interference effect to the DM production cross section,
 $\{ \sigma_{DM jj}(\phi=-\pi/4) - \sigma_{DM jj}(\phi=+\pi/4) \} /\sigma_{DM jj}(\phi=-\pi/4)$.
Unfortunately, the $Z$ + jets background is not significantly diminished by this requirement, 
 resulting in a significance factor that is smaller for the $DiJ$-$d$ cut than for the $DiJ$-$b$ cut.

\begin{table}
\begin{center}
\begin{tabular}{|c|c|c|c|c|c|} \hline
Cut           & $\sigma_{Z jj}$  & $\sigma_{DM jj}(\phi={\pi\over4})$ & $\sigma_{DM jj}(\phi=-{\pi\over4})$ &  $\frac{ \sigma_{DM jj}(-\pi/4)-\sigma_{DM jj}(\pi/4) }{ \sigma_{DM jj}(-\pi/4) }$ & $\frac{ \sigma_{DM jj}(-\pi/4)-\sigma_{DM jj}(\pi/4) }{ \sqrt{\sigma_{Z jj}+\sigma_{DM jj}(-\pi/4)} }$ \\ \hline
$DiJ$-$a$     & 49.8              & 9.246                        &  9.595                       &   0.0364                            & 0.0453   \\ \hline
$DiJ$-$b$     & 20.1             & 4.094                        &  4.355                        &   0.0599                          & 0.0528   \\ \hline
$DiJ$-$c$     & 8.34            & 1.868                        &  1.999                         &   0.0655                            &  0.0407         \\ \hline
$DiJ$-$d$     & 13.7             & 1.847                        &  2.012                        &   0.0820                          & 0.0416     \\ \hline          
\end{tabular} 
\end{center}
\caption{
Cross sections (in fb), the dimensionless ratio of the interference effect to the DM production cross section, and the significance factor (fb$^{1/2}$) 
 for DM production with \mbox{$m_{\chi}=10$~GeV}, $\Lambda=800$ GeV and $\phi=\pm \pi/4$, for four different cuts.
}
\end{table}

\begin{figure}[tbp]
  \begin{center}
   \includegraphics[width=90mm]{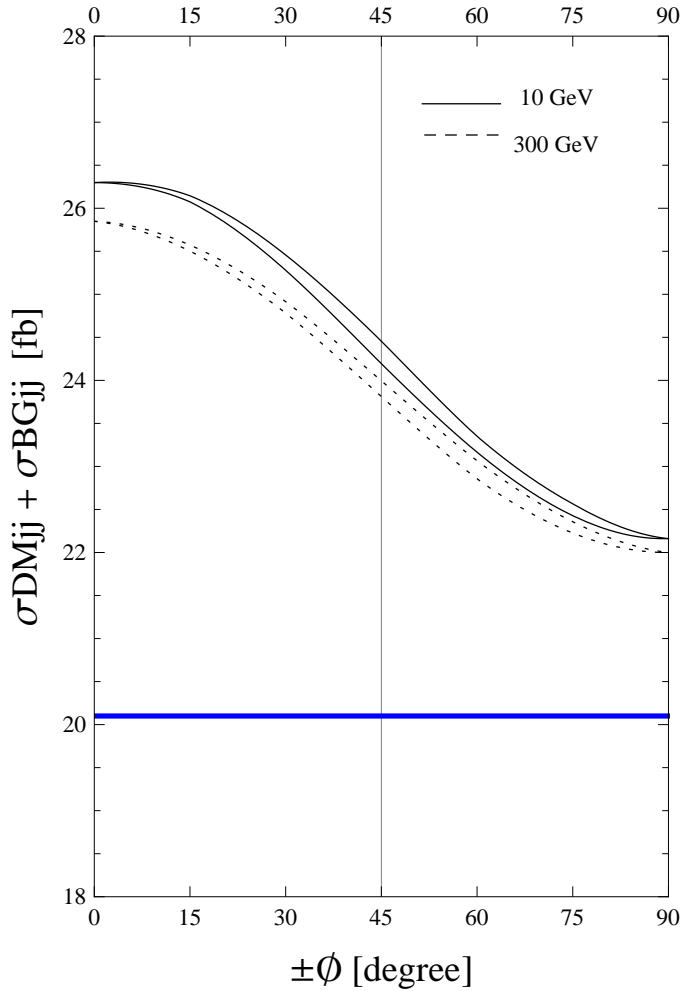}
  \end{center}
 \caption{The $Z$ + jets background cross section and the DM production cross section
  for \mbox{$\Lambda=800$~GeV} that satisfy the $DiJ$-$b$ selection cut,
  as a function of $\phi$ for the couplings
  $(g_Q, g_U, g_D)=(0, \cos \phi, \sin \phi)$.
 The horizontal solid line corresponds to the background cross section.
 The solid and dotted curves correspond to 
  the background + signal cross sections for $m_{\chi}=10$ GeV and $300$ GeV, respectively.
 For each pair of curves, the upper one corresponds to $\phi < 0$
  and the lower one to $\phi > 0$.
  }
  \label{fig5}
\end{figure}

\begin{figure}[tbp]
  \begin{center}
   \includegraphics[width=85mm]{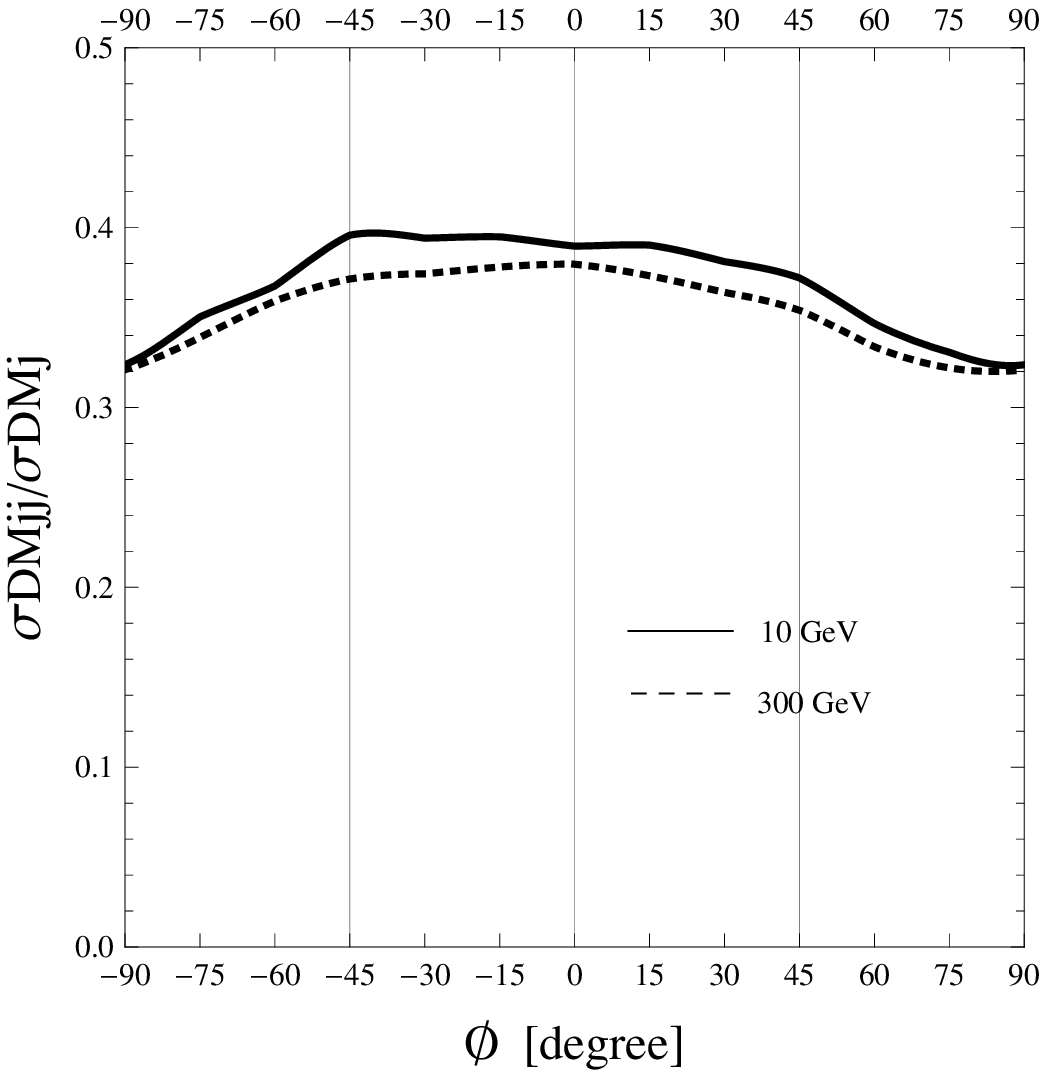}
  \end{center}
\caption{Ratio of the DM production cross section in the di-jet channel with the $DiJ$-$b$ cut
 to the cross section in the mono-jet channel with the $MonoJ14TeV$ cut for two values of  $m_\chi$.
  }
  \label{fig6}
\end{figure}

In Fig.~\ref{fig5} and Table~6, we show cross sections for the $Z$ + jets background and the DM production
 process for $\Lambda=800$ GeV and $m_{\chi}=10$, 300~GeV that satisfy the $DiJ$-$b$ selection cut.
The numbers in the table are in units of $(800\, {\rm GeV} / \Lambda)^{4}$~fb.
Finally in Fig.~\ref{fig6}, we show the ratio of the  DM signal cross section in the di-jet channel with the $DiJ$-$b$ cut to 
the cross section in the mono-jet channel with the $MonoJ14TeV$ cut, as a function of $\phi$.
\begin{table}
\begin{center}
\begin{tabular}{|c|c|c|} \hline
$m_{\chi}$ & $\phi= +\pi/4$ & $\phi= -\pi/4$ \\ \hline
10 GeV  & 4.094  &  4.355  \\ \hline
300 GeV & 3.712  &  3.895 \\ \hline
\end{tabular}
\end{center}
\caption{DM production cross sections (in units of $(800\, {\rm GeV} / \Lambda)^{4}$~fb) for $\sqrt{s}=14$ TeV that 
 satisfy the $DiJ$-$b$ selection cut.
}
\end{table}

\subsection{Discussion of the Di-jet Channel}

\ \ \ From Fig.~\ref{fig5} and Table~6, we confirm that
 the Feynman diagrams of the subprocess, $u_R d_R \rightarrow \chi \bar{\chi} \ u_R d_R$, interfere destructively
 if $g_{U}$ and $g_{D}$ have the same sign,
 and constructively if they have opposite signs.
This is in accordance with our expectation in Section 5.1
 based on the null radiation zone theorem \cite{null zone}.
Table~6 also indicates that
 the effect of the interference is diminished for larger DM masses.
This is because for heavier DM,
 the physical region deviates more
 from the null radiation zone given by Eq.~(\ref{nrzt}).


The deviation of Fig.~\ref{fig6} from a left-right symmetric form
 is a consequence of the interference.
Apart from the left-right asymmetry,
 Fig.~\ref{fig6} shows that the cross section ratio is smaller for $\phi \simeq \pm \pi/2$
 than for $\phi \simeq 0$.
This is due to the difference in the parton distribution functions of the up and down quarks;
 since the momentum distribution of the down quark in a proton
 leans towards a smaller momentum region compared to that of the up quark,
 events involving only down quarks have relatively less parton center-of-mass energy
 and are more likely to be rejected by 
 the high cut on $M_T(jj; E_{{\rm T}}\hspace{-1.10em}/ \ \ )$.

Once it has been experimentally established that $\vert g_{U} \vert \simeq \vert g_{D} \vert \ (\vert \phi \vert \simeq \pi/4)$,
 and that \mbox{$m_\chi \sim 10$~GeV},
 the sign of $g_U g_D$ can be determined
 by measuring the following DM cross section ratio 
 and comparing it with the prediction of Fig.~\ref{fig6}:
\begin{eqnarray}
\frac{ \sigma_{DMjj} }{ \sigma_{DMj} } &=& 
 \frac{ (N^{jj}_{obs} - N^{jj}_{SM})(DiJ{\rm \mathchar`-}b) }
 { (N^{j}_{obs} - N^{j}_{SM})(MonoJ14TeV) } \ ,
\end{eqnarray}
 where $N^{jj}_{obs}$ is the observed number of di-jet events
 and $N^{jj}_{SM}$ is the SM expectation for di-jet events.

We estimate the integrated luminosity needed to determine the sign of $g_{U} g_{D}$
 for \mbox{$\Lambda=800$~GeV}, $m_{\chi}=10$~GeV and $\vert \phi \vert=\pi/4$.
To distinguish $\phi=-\pi/4$ from
 $\phi=+\pi/4$ at the $2\sigma$ level, 
 the dijet-to-monojet cross section ratio must be measured with 3\% accuracy since
  Table~5 shows that the difference between the 2 cases is about 6\%.
The statistical uncertainty of $\sigma_{DM jj}/\sigma_{DMj}$ is given by
\begin{eqnarray}
\Delta \left( \frac{ \sigma_{DM jj} }{ \sigma_{DMj} } \right) &=& 
\left( \frac{ \sigma_{DM jj} }{ \sigma_{DMj} } \right) \ \sqrt{
\frac{ N^{j}_{obs}+N^{j}_{SM} }{ (N^{j}_{obs} - N^{j}_{SM})^{2} }
\ + \ \frac{ N^{jj}_{obs}+N^{jj}_{SM} }{ (N^{jj}_{obs} - N^{jj}_{SM})^{2} }
} \ .
\end{eqnarray}
Inserting $N_{SM}^{jj} = L \times 20.1$~fb and
 $N_{obs}^{jj}-N_{SM}^{jj} = L \times 4.355$~fb from Table~5,
 and $N_{SM}^{j} = L \times (12.8+2.4+2.4)$~fb from Eqs.~(\ref{zjets}-\ref{wjets}) 
 and $N_{obs}^{j}-N_{SM}^{j} = L \times 11.0$ fb from Table~2,
we find that to achieve a 3\% measurement of the dijet-to-monojet cross section ratio,
$L \ \simeq \ 3000 \ {\rm fb}^{-1}$ is needed.
\\

\section{Dark matter with suppressed scattering on xenon}

\ \ \ We now derive constraints on the contact interaction scale $\Lambda$ of our model Eq.~(\ref{lag}) for couplings that evade bounds from the xenon-based LUX direct detection experiment~\cite{lux}.
At present, LUX imposes a strong constraint on
 the spin-independent cross section for DM-nucleus elastic scattering.
Since the DM candidate of our model is a Dirac fermion, LUX imposes
 a severe lower bound on the contact interaction scale $\Lambda$ (for $g_Q,g_U,g_D$ normalized as $g_Q^2+g_U^2+g_D^2=1$)
 for general values of the ratios of $g_Q, g_U$ and $g_D$.
For example, for $g_Q=g_U=0$ and $g_D=1$, the lower bound on $\Lambda$ from the LUX data
 far exceeds 10 TeV for 10~GeV$\lesssim m_{\chi} \lesssim$500~GeV.
However, for $g_Q=0$ and $g_D/g_U=-0.89$, the bound becomes significantly weaker
 because contributions from the DM-up quark coupling and DM-down quark coupling mostly cancel for the proton and neutron content of Xe isotopes.
In this section, we take $g_Q=0$ and $g_D/g_U=-0.89$ ,
 so that constraints from the LUX and XENON~100~\cite{xenon} experiments are weakened.
With this choice of couplings, we obtain LHC bounds on the contact interation scale by normalizing the coupling constants as $g_U^2+g_D^2=1$.
We employ the latest results of the CMS collaboration from mono-jet searches with 19.5 fb$^{-1}$ of data at the 8~TeV 
LHC~\cite{cms-monoj}, 
 and mono-photon searches redone with 5.0 fb$^{-1}$ of data at the 7~TeV LHC~\cite{cms-monog}.
Bounds on $\Lambda$ derived from the di-jet channel are weaker than
those from the mono-jet channel.
\\

To derive bounds from the CMS mono-jet data,
 we calculate the DM signal cross section in the following way.
We generate $p p \rightarrow \chi \bar{\chi}$ + 1, 2 jets events with the $pp$ center-of mass energy of 8~TeV
 at the matrix element level with the precuts \met $> 200$~GeV and $k_{T} > 140$~GeV for the jets.
Parton showering is simulated with a 200~GeV matching scale,
 and detector effects are simulated. 
We implement the following selection cut, \textit{LATEST-MONOJ}, that mimics the one used in Ref.~\cite{cms-monoj}:
\begin{itemize}
 \item Require \met $>400$~GeV.
 \item Highest $p_T$ jet should satisfy $\vert \eta_1 \vert < 2.4$ and $p_{T1} > 110$~GeV.
 \item If a second highest $p_T$ jet with $\vert \eta_2 \vert < 4.5$ exists, 
 the azimuthal angle between this jet and the highest $p_T$ jet should satisfy $\Delta \phi(j_1, \, j_2) < 2.5$.
 \item Event is vetoed if three or more jets satisfy $\vert \eta \vert < 4.5$ and $p_T > 30$ GeV.
 \item Event is vetoed if there is an electron that satisfies $\vert \eta_{e} \vert < 2.47$ and $p_{T e} > 10$~GeV.
 \item Event is vetoed if there is a muon that satisfies $\vert \eta_{\mu} \vert < 2.4$ and $p_{T \mu} > 10$~GeV.

\end{itemize}

To calibrate the DM signal cross section calculated above,
 we calculate the cross section for the $Z(\nu \nu)$ + jets background (that has similar kinematics to the DM signal process), in the same way as the DM signal cross section.
We compare this cross section with the $Z(\nu \nu)$ + jets background cross section evaluated by CMS collaboration \cite{cms-monoj}
 which is (2596/19.5) fb,
 and then rescale the DM signal cross section by their ratio.

Using Eq.~(\ref{chi2}), we obtain the 90\%~C.L. lower bound on $\Lambda$ with 19.5 fb$^{-1}$ of data
 for each value of the DM mass $m_{\chi}$.
Here $N_{DM}(\Lambda)$, which scales as $1/\Lambda^4$, is the number of DM signal events that is based on the DM signal cross section derived and calibrated above.
The number of observed events $N_{obs}$, the number of expected SM events $N_{SM}$ and its uncertainty, $\sigma_{SM}$, are reported by the CMS collaboration to be 3677, 3663, and 196, respectively~\cite{cms-monoj}.
\\

A bound from CMS mono-photon data is derived in a similar fashion.
To calculate the DM signal cross section
we generate $p p \rightarrow \chi \bar{\chi}$ +  $\gamma$, $\chi \bar{\chi}$ + $\gamma$ + 1 jet events
 with the $pp$ center-of mass energy of 7 TeV
 at the matrix element level with the precuts $p_{T\gamma} > 60$~GeV for the photon and $k_{T} > 60$~GeV for the jet,
 and simulate parton showering with a matching scale of $84$~GeV.
Finally, detector simulations are performed.
We implement the following selection cut, \textit{LATEST-MONOG}, that mimics the one used in Ref.~\cite{cms-monog}:
\begin{itemize}
 \item Require a photon with $\vert \eta_{\gamma} \vert < 1.44$ and $p_{T \gamma} > 145$~GeV.
 \item Require \met $ > 130$~GeV.
 \item Event is vetoed if there is a jet with $\vert \eta_{j} \vert < 3.0$, $p_{T j} > 40$~GeV and $\Delta R(j,\gamma) < 0.5$.
 \item Event is vetoed if there is an isolated lepton with $p_{T l} > 20$~GeV and $\Delta R(l,\gamma) > 0.04$.
\end{itemize}

The DM signal cross section derived above is calibrated in the same way as for the mono-jet search.
We calculate the cross section for the $Z(\nu \nu)$ + $\gamma$ background in the same way as the DM signal,
 compare this cross section with the $Z(\nu \nu)$ + $\gamma$ background cross section evaluated by the CMS collaboration~\cite{cms-monog}
 which is (45.3/5.0) fb,
 and then rescale the DM signal cross section by their ratio.

We derive the 90\% C.L. lower bound on $\Lambda$ with 5.0 fb$^{-1}$ of data using Eq.~(\ref{chi2}).
Here $N_{DM}(\Lambda)$ is the number of DM signal events that is based on the DM signal cross section derived and calibrated above, and
$N_{obs}, N_{SM}$ and $\sigma_{SM}$
 are 75, 75.1 and 9.4, respectively~\cite{cms-monog}.
\\

\begin{figure}[tbp]
  \begin{center}
   \includegraphics[width=100mm]{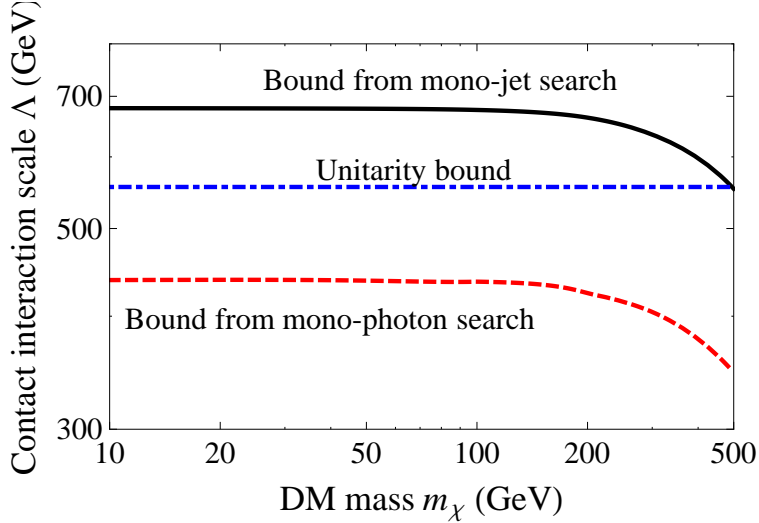}
  \end{center}
\caption{90\%~C.L. lower bounds on the contact interaction scale $\Lambda$
  for $g_Q=0$, $g_U=0.75$, $g_D=-0.66$,
  derived from the latest CMS mono-jet~\cite{cms-monoj} and
  mono-photon~\cite{cms-monog}  data. Note that with this choice of couplings, the DM scattering cross section in xenon-based detectors like LUX is significantly suppressed. 
 Also shown is the lower bound required by perturbative unitarity of the process $u_R \bar{u}_R \rightarrow \chi \bar{\chi}$
  with a parton center-of-mass energy of $\sqrt{\hat{s}}=3$ TeV.
  }
  \label{bounds}
\end{figure}

In Fig.~\ref{bounds}, we display 90\%~C.L. lower bounds on the contact interaction scale $\Lambda$ 
 for $g_Q=0$, $g_U=0.75$, $g_D=-0.66$ (which corresponds to $g_D/g_U=-0.89$, $g_U^2+g_D^2=1$) for various $m_{\chi}$.
For comparison, we also show the lower bound on $\Lambda$ required by 
 absolute perturbative unitarity of the process $u_R \bar{u}_R \rightarrow \chi \bar{\chi}$ for $g_U=0.75$ 
 with a parton center-of-mass energy, $\sqrt{\hat{s}}=3$ TeV.
(The unitarity bound for the process $d_R \bar{d}_R \rightarrow \chi \bar{\chi}$ for $g_D=-0.66$ is weaker.)
Details of absolute perturbative unitarity are explained in the Appendix.

\section{Mediator effects}

\ \ \ We now consider how the presence of a mediator field
 that contributes to resonant DM production changes our results.
We introduce a vector field, $V_{\mu}$, with TeV scale mass
 that couples to SM quarks and DM through the following Lagrangian:
\begin{eqnarray}
{\cal L}_{mediator} &=& \left(\bar{g}_U \bar{u}_R \gamma_{\mu} u_R \ + \ 
\bar{g}_D\bar{d}_R \gamma_{\mu} d_R \ + \ 
\bar{g}_{\chi} \bar{\chi} \gamma_{\mu} \chi \right)V^{\mu} 
\ - \ \frac{1}{2} M_V^2 V^{\mu} V_{\mu} \ . \label{med lag}
\end{eqnarray}
 For large $M_V$, we recover the interaction term of Eq.~(\ref{lag}) (with $g_Q=0$)
with the correspondence:
\begin{eqnarray}
\frac{\bar{g}_U \bar{g}_{\chi}}{M_V^2} &=& \frac{g_U}{\Lambda^2} \ ,
\ \ \ \ \ \ \ \ \ \ \frac{\bar{g}_D \bar{g}_{\chi}}{M_V^2} \ = \ \frac{g_D}{\Lambda^2} \ .
\label{map}
\end{eqnarray}
If $V_{\mu}$ has a mass of several TeV,
 the DM production cross section at the LHC
 is enhanced by the $V_\mu$ resonance.
We calculate how the cross sections for
 mono-jet, mono-photon and di-jet events depend on $M_V$.

For our numerical study,
 we require the coupling constants $\bar{g}_U$, $\bar{g}_D$ and $\bar{g}_{\chi}$ to satisfy
\begin{eqnarray}
\frac{\bar{g}_U \bar{g}_{\chi}}{M_V^2} &=& \frac{\bar{g}_D \bar{g}_{\chi}}{M_V^2} 
\ = \ \frac{1}{\sqrt{2}} \frac{1}{(800 \ {\rm GeV})^2} \ , \label{med constraint}
\end{eqnarray}
 so that in the absence of the $V_\mu$ resonance, we obtain a contact interaction
 with $\Lambda=800$~GeV and $g_U=g_D=1/\sqrt{2}$. This allows us to quantify the
 effect of the mediator.
We select the following $V_\mu$ masses:
\begin{align}
 & M_V \ = \ 2.0 \ {\rm TeV}, \ \ \ 2.4 \ {\rm TeV}, \ \ \ 3.2 \ {\rm TeV}\,.
\end{align}
 We do not consider values of $M_V \lesssim 1.6$~TeV because these 
 are excluded by the 8 TeV LHC data~\cite{8tev monojet} for coupling constants satisfying Eq.~(\ref{med constraint}).
We choose values of $\bar{g}_U(=\bar{g}_D)$ and $\bar{g}_{\chi}$ 
 so that the decay width of $V_{\mu}$ is minimized
 and the resonant production of $\chi$ is maximally enhanced.
Since the width of $V_{\mu}$ is given by
\begin{eqnarray}
\Gamma_V &=& \frac{ 3 \bar{g}_U^2 + 3 \bar{g}_D^2 + 2 \bar{g}_{\chi}^2 }{24 \pi} M_V \ ,
\end{eqnarray}
 the width is minimized if $\bar{g}_\chi=\sqrt{3}\bar{g}_U$ 
 for $\bar{g}_U = \bar{g}_D$.
\\

We evaluate the cross sections for mono-jet events at the LHC with $\sqrt{s}=8$~TeV and 14~TeV,
 mono-photon events at 14~TeV and di-jet events at 14~TeV.
For mono-jet events at 8 TeV,
 we implement a cut that resembles the \textit{SR4} selection cut used by the ATLAS collaboration~\cite{8tev monojet},
 and compare the cross sections in our benchmark models with the ATLAS data~\cite{8tev monojet}.
Our cut, \textit{MonoJ8TeV}, is as follows:
\begin{itemize}
 \item Require one jet with $\vert \eta \vert < 2.0$ and $p_{T} > 500$~GeV.
 \item Require \met $> 500$~GeV.
 \item Event is vetoed if three or more jets satisfy $\vert \eta \vert < 4.5$ and $p_{T} > 30$~GeV.
 \item Event is vetoed if there is an electron that satisfies $\vert \eta_{e} \vert < 2.47$ and $p_{T e} > 20$~GeV.
 \item Event is vetoed if there is a muon that satisfies $\vert \eta_{\mu} \vert < 2.4$ and $p_{T \mu} > 7$~GeV.
\end{itemize}
We simulate the signal $p p \rightarrow \chi \bar{\chi} + jets$ events 
 using the procedure described in Section 3,
 with the precuts $k_T>180$~GeV and \met$>300$~GeV, and a 250~GeV matching scale,
 which we have confirmed to be appropriate.
We analogously simulate $p p \rightarrow Z(\nu \nu) + jets$ events and compare the cross section after the \textit{MonoJ8TeV} cut 
 with the one after the \textit{SR4} cut estimated by the ATLAS collaboration~\cite{8tev monojet}.
We find that the two cross sections match,
 so we do not need to rescale our signal cross section.
Using the number of observed events ($N_{obs}$) and the background prediction ($N_{SM} \pm \sigma_{SM}$)
 reported in Ref.~\cite{8tev monojet}, and our estimate of $N_{DM}$ for a particular $M_V$,
 we employ the criterion of Eq.~(\ref{chi2}) to assess if the value of $M_V$ is allowed at 90\% C.L. We find $\chi^2(M_V=2\ \rm{TeV})=2.15$
 and $\chi^2(M_V=2.4\ \rm{TeV})=1.54$. No signal events are produced for $M_V=3.2$~TeV at the 8~TeV LHC. Thus $M_V>2$~TeV is unconstrained
 by current LHC data.
%
%

We implement the \textit{MonoJ14TeV},  \textit{MonoG14TeV-a} and \textit{DiJ-b}  cuts for mono-jet, mono-photon and di-jet events at the 14 TeV LHC, 
respectively.
The cross sections are displayed in Table 7. We see that the cross sections at the 14 TeV LHC are significantly enhanced
 for a $\sim 2$~TeV mass mediator. The cross sections for a 3.2~TeV mediator are reduced because the width of the mediator 
 satisfying Eq.~(\ref{med constraint}) is as large as its mass.

\begin{table}
\begin{center}
\begin{tabular}{|c|c|c|c|c|} \hline
              & $M_V=2.0$ TeV & $M_V=2.4$ TeV & $M_V=3.2$ TeV & Contact interaction \\ \hline
$\sigma(MonoJ8TeV)$& 6.33    & 3.72    & -       & -                      \\ \hline
$\sigma(MonoJ14TeV)$ & 27.1    & 17.5    & 7.48    & 11.0                   \\ \hline
$\sigma(MonoG14TeV$-$a)$ & 12.4 & 8.15   & 3.53    & 5.49                \\ \hline
$\sigma(DiJ$-$b)$     & 10 & 6.0 & 2.6          & 4.09                  \\ \hline
\end{tabular}
\end{center}
\caption{
Cross sections (in fb)
 for mono-jet events at the 8 TeV LHC, and mono-jet, mono-photon and di-jet events at the 14 TeV LHC (for the cuts specified), for models with a mediator of mass
 $M_V$ and minimum width, and for a contact interaction.
}
\end{table}


\section{Discussion and Conclusion}

\ \ \ We explored ways of observing the isospin violating nature of a dark matter particle
 at the LHC.
We pursued the possibility that DM interacts with SM quarks via couplings that violate isospin
 by considering a toy model.
We adopted a Dirac fermion DM particle that couples to
 up and down quarks through contact interactions with different strengths.
The cross section ratio
 of the DM production associated with a hard jet and that with a photon
 reflects the absolute value of the ratio of the DM couplings to up and down quarks
 because they have different electric charges.
We showed that this ratio can be measured at the LHC 
 by observing and comparing \mbox{mono-photon + \met} events with mono-jet + \met events.

The relative sign of the DM couplings to up and down quarks
 can be studied only by comparing the cross section of DM production associated with two hard jets
 to that with one hard jet.
This is because the subprocess, $u_R d_R \rightarrow u_R d_R \chi \bar{\chi}$, that contributes
 to \mbox{di-jet + \met} events,
 contains an interference term that is proportional to $g_{U}g_{D}$.
In order to extract the interference term,
 we implemented a very hard cut on the di-jet cluster transverse mass,
 $M_T(jj; E_{{\rm T}}\hspace{-1.10em}/ \ \ )$,
 which selects quark-quark collision events over quark-gluon collision backgrounds.
We showed that the effect of the interference does in fact appear in the cross section of 
 di-jet + \met events
 with the sign expected from the null radiation zone theorem~\cite{null zone},
 and that it is possible to determine the relative sign of the couplings 
 by observing and comparing di-jet + \met events with mono-jet + \met events.
\\

In Sections 3, 4, 5 and 6,
 we studied the scenario in which the DM couples to quarks through
 contact interactions of dimension-6.
In the Appendix we show that in this scenario,
 the perturbative unitarity bound for the DM production process
 is violated when the DM invariant mass is large.
We therefore assumed that the DM production cross section is constant 
 when the DM invariant mass is larger than the absolute perturbative unitarity bound of
 Eq.~(\ref{uni bound for lag}).
However, the DM production cross section in $pp$ collisions at the 14~TeV LHC
 is only slightly affected by this modification
 because the number of events in which the DM invariant mass is larger than the unitarity bound
 is exponentially suppressed by the parton distribution functions of the proton.

In Section 7, we considered a model with a mediator field,
 and studied how the cross sections change with the mass of the mediator.
We showed that the cross sections for mono-jet, mono-photon and di-jet events at the 14~TeV LHC
 can be enhanced by the mediator resonance
 without conflicting with mono-jet searches
 at the 8 TeV LHC~\cite{8tev monojet}.
Prospects for discovery improve if the DM and quarks couple through a $\sim 2$ TeV
 mediator, instead of via contact interactions.
\\

Throughout, we focused on the case with $g_{Q}=0$.
If $g_{Q} \neq 0$, the DM cross section ratios in Figs.~\ref{fig3} and~\ref{fig6}, will be modified.
For Fig.~\ref{fig3}, the angle $\phi$ will be replaced by
\begin{eqnarray*}
\phi_{eff} &=& \tan^{-1} \ \left( \ \sqrt{g_{D}^{2}+g_{Q}^{2}} \ / \ \sqrt{g_{U}^{2}+g_{Q}^{2}} \ \right) \ .
\end{eqnarray*}
On the other hand, Fig.~\ref{fig6} will be modified in a complicated way.
Note that the ratio of the di-jet and mono-jet signal cross sections
 can be written in the following way when $g_{Q}=0$:
\begin{eqnarray}
\sigma_{jj}/\sigma_{j} &=& 
\frac{ C g_{U}^{2} + D g_{U}g_{D} + E g_{D}^{2} }{ A g_{U}^{2} + B g_{D}^{2} }\,,
\end{eqnarray}
 where $A,B,C,D,E$ are numerical factors with $A,B,C,E>0$ and $D<0$ as we saw in Section 5.
For $g_{Q}\neq 0$, the expression above becomes:
\begin{eqnarray}
\sigma_{jj}/\sigma_{j} &=& 
\frac{ C (g_{U}^{2}+g_{Q}^{2}) + D (g_{U}g_{D}+g_{Q}g_{U}+g_{Q}g_{D}) + E (g_{D}^{2}+g_{Q}^{2}) }
{ A (g_{U}^{2}+g_{Q}^{2}) + B (g_{D}^{2}+g_{Q}^{2}) } \ .
\end{eqnarray}
Our analysis can be generalized to the entire parameter space of ($g_{Q}, g_{U}, g_{D}$).
\\

Although this work is motivated by isospin violating dark matter,
 our methods are applicable to other new physics models
 for which one studies the ratio of the new physics couplings to up and down quarks.
The di-jet channel is challenging because it is necessary to suppress large contributions from
 gluon-quark interactions.
However, it is the only channel that is sensitive to the relative sign of the up and down
 quark couplings, and hence should be studied seriously 
 once new physics that couples to quarks is discovered.
We believe that our exploratory studies will  be useful
 to probe new physics properties at the LHC.
\\

\section*{Acknowledgments}

\ \ \ D.M. thanks the KEK Theory Center, where this work was initiated, for its support and hospitality. He 
also thanks the Center for Theoretical Underground Physics and Related Areas (CETUP* 2012) in South Dakota 
for its support and hospitality during the completion of this work.
T.Y. thanks the organizers and the lecturers of ``KIAS School on MadGraph for LHC Physics Simulation"
 (24-29 October, 2011) \cite{kias}, where physics simulation tools used in this study
 were introduced. This work was supported in part by Grant-in-Aids for Scientific Research
 (Nos. 23-3599, 20340064 and 23104006) from
 the Japan Society for the Promotion of Science, and by US DOE
grants DE-FG02-04ER41308 and DE-FG02-13ER42024, and US NSF grant PHY-0544278.
\\

\section*{Appendix: \ Unitarity bounds for contact interactions}

\subsection*{Unitarity of scattering amplitudes}

\ \ \ We begin with a review of unitarity constraints on general scattering amplitudes.
The S-matrix can be written as
\begin{eqnarray}
S &\equiv& 1 +  i T \,, \label{uni}
\end{eqnarray}
 where $T$ gives the transition amplitudes.
In terms of $T$, the unitarity of the S-matrix, $S^{\dagger} S = 1$, is
\begin{eqnarray}
-i (T-T^{\dagger}) &=& T^{\dagger} T \,. \label{uni1}
\end{eqnarray}
For an elastic scattering process $A + B \rightarrow A + B$,
 the above equation gives
\begin{eqnarray}
& & -i (\langle \vec{p}_A \vec{p}_B \vert T \vert \vec{k}_A \vec{k}_B \rangle 
 -  \langle \vec{p}_A \vec{p}_B \vert T^{\dagger} \vert \vec{k}_A \vec{k}_B \rangle) \nonumber \\
&=& \langle \vec{p}_A \vec{p}_B \vert T^{\dagger} T \vert \vec{k}_A \vec{k}_B \rangle \ \nonumber \\
&=& \sum_{\{q_i\}} \left( \ \prod_i \int \frac{ {\rm d}^3 \vec{q}_i }{ (2\pi)^3 2 q^0_i } \ \right) \
            \langle \vec{p}_A \vec{p}_B \vert T^{\dagger} \vert \{\vec{q}_i\} \rangle
            \langle \{\vec{q}_i\} \vert T \vert \vec{k}_A \vec{k}_B \rangle \ , \label{uni2}
\end{eqnarray}
 where $\vert \{\vec{q}_i\} \rangle$ constitute a complete set of states.
The scattering matrix element, ${\cal M}$, for the scattering process,
\begin{eqnarray}
A \ + \ B &\rightarrow& 1 \ + \ 2 \ + \ ... \ + \ F\,,
\end{eqnarray}
 can be expressed as
\begin{eqnarray}
& & i {\cal M} (\vec{k}_A \vec{k}_B \rightarrow \vec{p}_1 \vec{p}_2 ... \vec{p}_F)  
(2\pi)^4 \delta^4 (k_A + k_B - p_1 - p_2 ... - p_F) 
\equiv \langle \vec{p}_1 \vec{p}_2 ... \vec{p}_F \vert i T \vert \vec{k}_A \vec{k}_B \rangle \ ,
\end{eqnarray}
 where we denote the three- and four-momenta of the initial-state particles
 by $\vec{k}_A, \ \vec{k}_B$ and $k_A, \ k_B$, respectively,
 and those of the final-state particles by
 by $\vec{p}_1, \ \vec{p}_2, \ ..., \ \vec{p}_F$ and $p_1, \ p_2, \ ..., \ p_F$.
In terms of matrix elements, the unitarity relation Eq.~(\ref{uni2}) reads
\begin{eqnarray}
& & -i \left\{ \ {\cal M}(\vec{k}_A \vec{k}_B \rightarrow \vec{p}_A \vec{p}_B) \ - \ 
{\cal M}(\vec{p}_A \vec{p}_B \rightarrow \vec{k}_A \vec{k}_B)^* \ \right\} \nonumber \\
&=& \sum_{\{q_i\}} \left( \ \prod_i \int \frac{ {\rm d}^3 \vec{q}_i }{ (2\pi)^3 2 q^0_i } \ \right) \
{\cal M}(\vec{p}_A \vec{p}_B \rightarrow \{\vec{q}_i\})^* {\cal M}(\vec{k}_A \vec{k}_B \rightarrow \{\vec{q}_i\}) \ 
(2\pi)^4 \delta^4 \left( k_A + k_B - \sum_i q_i \right) \,. 
\nonumber \\
\label{uni3}
\end{eqnarray}
The unitarity bounds for both elastic and inelastic scattering processes
 follow from Eq.~(\ref{uni3}).
\\

\subsection*{Absolute perturbative unitarity bound for DM production}

\ \ \ We are interested in the perturbative unitarity bound for the DM production process
 arising from the contact interaction terms in Eq.~(\ref{lag}),
\begin{eqnarray}
q_R^{c1} \ \bar{q}_{R \, c1} &\rightarrow& \chi \bar{\chi} \ \ (q=u,d) \,, \label{dm production}
\end{eqnarray}
 where $c1$ is a color index.
In this subsection, we discuss the \textit{absolute} unitarity bound for the above process 
 which holds regardless of the details of other processes,
 such as $q \bar{q} \rightarrow q^{\prime} \bar{q}^{\prime}$ and $\chi \bar{\chi} \rightarrow \chi \bar{\chi}$.
For simplicity, we assume that the quarks and the DM are massless.
If the DM has a non-negligible mass, the unitarity bound weakens.

Consider the elastic scattering processes, $q_R^{c1} \bar{q}_{R \, c1} \rightarrow q_R^{c2} \bar{q}_{R \, c2} \ (q=u,d)$,
 whose matrix elements we denote by ${\cal M}_{q q}$.
Also consider inelastic scattering processes, $q_R^{c1} \bar{q}_{R \, c1} \rightarrow \chi_R \bar{\chi}_R$
 and $q_R^{c1} \bar{q}_{R \, c1} \rightarrow \chi_L \bar{\chi}_L$,
 whose matrix elements we denote by ${\cal M}_{q \chi_R}$ and ${\cal M}_{q \chi_L}$,
 respectively.
From Eq.~(\ref{uni3}), we find
\begin{eqnarray}
& & -i \left\{ \ {\cal M}_{q q} (\vec{k}_A \vec{k}_B \rightarrow \vec{p}_A \vec{p}_B) \ - \ 
{\cal M}_{q q} (\vec{p}_A \vec{p}_B \rightarrow \vec{k}_A \vec{k}_B)^* \ \right\} \nonumber \\
&\geq& \int \frac{ {\rm d}^3 \vec{q}_1 }{ (2\pi)^3 2 q^0_1 } \int \frac{ {\rm d}^3 \vec{q}_2 }{ (2\pi)^3 2 q^0_2 } \
{\cal M}_{q q} (\vec{p}_A \vec{p}_B \rightarrow \vec{q}_1 \vec{q}_2)^* {\cal M}_{q q} (\vec{k}_A \vec{k}_B \rightarrow \vec{q}_1 \vec{q}_2) 
(2\pi)^4 \delta^4 \left( k_A + k_B - q_1 - q_2 \right) \nonumber \\
& & + \ \int \frac{ {\rm d}^3 \vec{q}_1 }{ (2\pi)^3 2 q^0_1 } \int \frac{ {\rm d}^3 \vec{q}_2 }{ (2\pi)^3 2 q^0_2 } \
{\cal M}_{q \chi_R} (\vec{p}_A \vec{p}_B \rightarrow \vec{q}_1 \vec{q}_2)^* {\cal M}_{q \chi_R} (\vec{k}_A \vec{k}_B \rightarrow \vec{q}_1 \vec{q}_2)  
(2\pi)^4 \delta^4 \left( k_A + k_B - q_1 - q_2 \right) \nonumber \\
& & + \ ({\cal M}_{q \chi_R}  \rightarrow  {\cal M}_{q \chi_L} ) \nonumber \\
&=& \frac{1}{ (2\pi)^2 4 q_1^0 q_2^0 } \int {\rm d}\Omega \ \vert \vec{q} \vert \frac{ q_1^0 q_2^0 }{ q_1^0+q_2^0 } \
{\cal M}_{q q} (\vec{p}_A \vec{p}_B \rightarrow \vec{q}_1 \vec{q}_2)^* {\cal M}_{q q} (\vec{k}_A \vec{k}_B \rightarrow \vec{q}_1 \vec{q}_2) \ 
\nonumber \\
& & + \ \frac{1}{ (2\pi)^2 4 q_1^0 q_2^0 } \int {\rm d}\Omega \ \vert \vec{q} \vert \frac{ q_1^0 q_2^0 }{ q_1^0+q_2^0 } \
{\cal M}_{q \chi_R} (\vec{p}_A \vec{p}_B \rightarrow \vec{q}_1 \vec{q}_2)^* {\cal M}_{q \chi_R} (\vec{k}_A \vec{k}_B \rightarrow \vec{q}_1 \vec{q}_2) 
\nonumber \\
& & + \ ({\cal M}_{q \chi_R}  \rightarrow  {\cal M}_{q \chi_L}) \,,
\label{uni for qdm}
\end{eqnarray}
 where $\Omega$ is the scattering solid angle in the center-of-mass frame,
 and ${\cal M}_{q \chi_R} \rightarrow {\cal M}_{q \chi_L}$
 is the term obtained by replacing ${\cal M}_{q \chi_R}$ 
 by ${\cal M}_{q \chi_L}$ in the previous term.
We make a partial wave expansion of ${\cal M}_{q q}$ as
\begin{eqnarray}
{\cal M}_{q q}(s, \cos \alpha) &\equiv& 16 \pi  \sum_{l=0}  a^{q q}_l(s) (2l+1) P_l(\cos \alpha) \,, 
\end{eqnarray}
 where $s$ is the center-of-mass energy and 
 $\alpha$ is the scattering angle in the center-of-mass frame.
Expanding the last two lines of Eq.~(\ref{uni for qdm}) in terms of $P_l(\cos \theta)$,
 where $\theta$ is the angle between $\vec{p}_A$ and $\vec{k}_A$ in the center-of-mass frame:
\begin{eqnarray}
& & \frac{1}{ (2\pi)^2 4 q_1^0 q_2^0 } \int {\rm d}\Omega \ \vert \vec{q} \vert \frac{ q_1^0 q_2^0 }{ q_1^0+q_2^0 } \
{\cal M}_{q \chi_R} (\vec{p}_A \vec{p}_B \rightarrow \vec{q}_1 \vec{q}_2)^* {\cal M}_{q \chi_R} (\vec{k}_A \vec{k}_B \rightarrow \vec{q}_1 \vec{q}_2)
\nonumber \\
&\equiv&
16 \pi  \sum_{l=0}  b^{q \chi_R}_l(s)  (2l+1)  P_l(\cos \theta) \,,
\nonumber \\
& & +\ ( R  \rightarrow  L  ) \,.
\end{eqnarray}
Then Eq.~(\ref{uni for qdm}) becomes
\begin{eqnarray}
2 {\rm Im}(a^{q q}_l) &\geq& 2 \vert a^{q q}_l \vert^2 \ + \ b^{q \chi_R}_l \ + \ b^{q \chi_L}_l \ ,
\ 
\nonumber \\
\end{eqnarray}
 which leads to the following bound on the matrix elements for the processes
 $\chi_R \bar{\chi}_R \rightarrow q_R \bar{q}_R$ and $\chi_L \bar{\chi}_L \rightarrow q_R \bar{q}_R$:
\begin{eqnarray}
b^{q \chi_R}_l \ + \ b^{q \chi_L}_l &\leq& \frac{1}{2} \ .
\label{uni-fin}
\end{eqnarray}
\\

To obtain the unitarity bound we evaluate $b^{\chi q}_l$'s perturbatively.
At the tree level, we have
\begin{eqnarray}
b^{q \chi_R}_0 &=& b^{q \chi_L}_0 \ = \ \frac{1}{192 \pi^2} \frac{s^2  g_q^2}{\Lambda^4} \ ,
\nonumber \\
b^{q \chi_R}_1 &=& b^{q \chi_L}_1 \ = \ \frac{1}{3 \cdot 192 \pi^2} \frac{s^2  g_q^2}{\Lambda^4} \ ,
\nonumber \\
b^{q \chi_R}_{l \geq 2} &=& b^{q \chi_L}_{l \geq 2} \ = \ 0 \ \ \ \ \ \ \ \ \ \ \ (q=u,d) \ .
\end{eqnarray}
We thus find that
 Eq.~(\ref{uni-fin}) is satisfied if the center-of-mass energy,
 $\sqrt{s} = M_{q\bar{q}} = M_{\chi \bar{\chi}}$,
 obeys the condition,
\begin{eqnarray}
\sqrt{s} &<& (48 \pi^2)^{1/4} \ \frac{\Lambda}{\sqrt{g_q}} \ = \ 
3.7 \ {\rm TeV}  \times \left( \frac{\Lambda}{800 \, {\rm GeV}} \right) \
\left( \frac{1}{g_q} \right)^{1/2} \label{bound on s}
\end{eqnarray}
 for $q=u,d$.

One can also derive a perturbative unitarity bound for DM production
 by considering the elastic scattering process, $\chi \bar{\chi} \rightarrow \chi \bar{\chi}$,
 instead of $q_R^{c1} \bar{q}_{R \, c1} \rightarrow q_R^{c2} \bar{q}_{R \, c2}$.
However, the resultant bound is weaker than Eq.~(\ref{bound on s}).
\\

\subsection*{Model-dependent perturbative unitarity bound for DM production}

\ \ \ In this subsection,
 we derive the perturbative unitarity bound for  DM production
 \textit{with specfic assumptions} on the underlying model.
Our assumptions are:

\begin{itemize}
 \item DM and quarks interact through contact terms in the Lagrangian,
\begin{eqnarray}
{\cal L}_{contact} &=& \frac{ \bar{g}_{\chi}^2 }{M_V^2} \ 
(\bar{\chi} \gamma^{\mu} \chi) \ (\bar{\chi} \gamma_{\mu} \chi)
\nonumber \\
&+& \frac{ \bar{g}_{\chi} \bar{g}_{U} }{M_V^2} \ (\bar{\chi} \gamma^{\mu} \chi) \ (\bar{u}_R \gamma_{\mu} u_R)
\ + \ \frac{ \bar{g}_{\chi} \bar{g}_{D} }{M_V^2} \ (\bar{\chi} \gamma^{\mu} \chi) \ (\bar{d}_R \gamma_{\mu} d_R)
\nonumber \\
&+& \frac{ \bar{g}_{U}^2 }{M_V^2} \ (\bar{u}_R \gamma^{\mu} u_R) \ (\bar{u}_R \gamma_{\mu} u_R)
\ + \ \frac{ \bar{g}_{U} \bar{g}_{D} }{M_V^2} \ (\bar{u}_R \gamma^{\mu} u_R) \ (\bar{d}_R \gamma_{\mu} d_R)
\nonumber \\
&+& \frac{ \bar{g}_{D}^2 }{M_V^2} \ (\bar{d}_R \gamma^{\mu} d_R) \ (\bar{d}_R \gamma_{\mu} d_R) \,,
\label{underlying lag}
\end{eqnarray}
 which naturally arises by integrating out a massive mediator vector field that couples to
 $u_R$, $d_R$ and $\chi$.
\item Interactions between DM particles and between DM and a quark
 are induced only by terms in Eq.~(\ref{underlying lag}).
\item Interactions between quarks
 are induced only by terms in Eq.~(\ref{underlying lag}),
 and by SM gauge interactions.
\end{itemize}
We note that the contact terms in Eq.~(\ref{underlying lag}) correspond to those in Eq.~(\ref{lag})
 according to Eq.~(\ref{map}).
%
%

We denote the matrix elements for scattering processes involving the DM and/or quarks,
 $A \bar{A} \rightarrow B \bar{B} \ \ (A,B = u_R^{c1}, \ d_R^{c2}, \ \chi_R, \ \chi_L)$,
 by ${\cal M}^{A B}$, and expand in partial waves,
\begin{eqnarray}
{\cal M}^{A B} (s, \cos \alpha) &=& 16 \pi  \sum_{l=0}  a^{A B}_l(s)  (2l+1) \ P_l(\cos \alpha) \,.
\end{eqnarray}
For each $l$, consider a matrix of the factors $a^{A B}_l(s)$, given by
\begin{eqnarray}
{\cal T}_l &\equiv& 
\left(
\begin{array}{cccc}
a^{u_R u_R}_l & a^{u_R d_R}_l & a^{u_R \chi_R}_l & a^{u_R \chi_L}_l \\
a^{d_R u_R}_l & a^{d_R d_R}_l & a^{d_R \chi_R}_l & a^{d_R \chi_L}_l \\
a^{\chi_R u_R}_l & a^{\chi_R d_R}_l & a^{\chi_R \chi_R}_l & a^{\chi_R \chi_L}_l \\
a^{\chi_L u_R}_l & a^{\chi_L d_R}_l & a^{\chi_L \chi_R}_l & a^{\chi_L \chi_L}_l
\end{array}
\right) \,,
\end{eqnarray}
 where the color indices in  $a^{A B}_l$'s are implicit. 
 We diagonalize ${\cal T}_l$'s by taking linear combinations of the initial and final states
 to find the eigenvalues $t_l^i$ $(i=1,2,...,8)$.
For each $l$, 
 the scattering processes involving the DM and/or quarks
 can be regarded as the elastic scatterings of the states in the new basis.
From Eq.~(\ref{uni3}), it follows that each $t_l^i$
 satisfies
\begin{eqnarray}
{\rm Im}(t_l^i) &\geq& \vert t_l^i \vert^2 \,,
\end{eqnarray}
 which leads to the bound,
\begin{eqnarray}
\vert t_l^i \vert &\leq& 1 \,. \label{md uni} 
\end{eqnarray}

The model-dependent unitarity bound is derived
 by evaluating $t_l^i$'s perturbatively,
 based on the assumptions above.
At the tree level, we have
\begin{eqnarray}
a^{u_R^{c1} u_R^{c2}}_0 &=& \frac{1}{16 \pi} \frac{\bar{g}_U^2}{M_V^2} s \ + \ \frac{1}{16 \pi} \frac{1}{3} g_s^2 \nonumber \\
a^{u_R^{c1} d_R^{c2}}_0 &=& a^{d_R^{c1} u_R^{c2}}_0 \ = \ \frac{1}{16 \pi} \frac{\bar{g}_U \bar{g}_D}{M_V^2} s \ + \ \frac{1}{16 \pi} \frac{1}{3} g_s^2
\nonumber \\
a^{d_R^{c1} d_R^{c2}}_0 &=& \frac{1}{16 \pi} \frac{\bar{g}_D^2}{M_V^2} s \ + \ \frac{1}{16 \pi} \frac{1}{3} g_s^2 \nonumber \\
a^{\chi_R u_R}_0 &=& a^{\chi_L u_R}_0 \ = \ 
a^{u_R \chi_R}_0 \ = \ a^{u_R \chi_L}_0 
\ = \ \frac{1}{16 \pi} \frac{\bar{g}_U \bar{g}_{\chi}}{M_V^2} s \nonumber \\
a^{\chi_R d_R}_0 &=& a^{\chi_L d_R}_0 \ = \ 
a^{d_R \chi_R}_0 \ = \ a^{d_R \chi_L}_0 
\ = \ \frac{1}{16 \pi} \frac{\bar{g}_D \bar{g}_{\chi}}{M_V^2} s \nonumber \\
a^{\chi_R \chi_R}_0 &=& a^{\chi_R \chi_L}_0 \ = \
a^{\chi_L \chi_R}_0 \ = \ a^{\chi_L \chi_L}_0
\ = \ \frac{1}{16 \pi} \frac{\bar{g}_{\chi}^2}{M_V^2} s \nonumber \\
a^{u_R^{c1} u_R^{c2}}_1 &=& \frac{1}{16 \pi} \frac{1}{3} \frac{\bar{g}_U^2}{M_V^2} s \ + \ \frac{1}{16 \pi} \frac{1}{3 \cdot 3} g_s^2 \nonumber \\
a^{u_R^{c1} d_R^{c2}}_1 &=& a^{d_R^{c1} u_R^{c2}}_1 \ = \ \frac{1}{16 \pi} \frac{1}{3} \frac{\bar{g}_U \bar{g}_D}{M_V^2} s \ + \ \frac{1}{16 \pi} \frac{1}{3 \cdot 3} g_s^2
\nonumber \\
a^{d_R^{c1} d_R^{c2}}_1 &=& \frac{1}{16 \pi} \frac{1}{3} \frac{\bar{g}_D^2}{M_V^2} s \ + \ \frac{1}{16 \pi} \frac{1}{3 \cdot 3} g_s^2 \nonumber \\
-a^{\chi_R u_R}_1 &=& a^{\chi_L u_R}_1 \ = \ 
-a^{u_R \chi_R}_1 \ = \ a^{u_R \chi_L}_1 
\ = \ \frac{1}{16 \pi} \frac{1}{3} \frac{\bar{g}_U \bar{g}_{\chi}}{M_V^2} s \nonumber \\
-a^{\chi_R d_R}_1 &=& a^{\chi_L d_R}_1 \ = \ 
-a^{d_R \chi_R}_1 \ = \ a^{d_R \chi_L}_1
\ = \ \frac{1}{16 \pi} \frac{1}{3} \frac{\bar{g}_D \bar{g}_{\chi}}{M_V^2} s \nonumber \\
-a^{\chi_R \chi_R}_1 &=& -a^{\chi_L \chi_L}_1 
\ = \ a^{\chi_R \chi_L}_1 \ = \ a^{\chi_L \chi_R}_1 
\ = \ \frac{1}{16 \pi} \frac{1}{3} \frac{\bar{g}_{\chi}^2}{M_V^2} s \nonumber \\
a^{A B}_{l\geq 2} &=& 0 \,,
\end{eqnarray}
 where $g_s$ is the SU(3)$_C$ gauge coupling.
We neglected electroweak gauge interactions, which give subdominant contributions
 relative to the SU(3)$_C$ gauge interaction.

As an example, for $M_V=800$~GeV, $\bar{g}_U=\bar{g}_D=1/\sqrt{2}$ and $\bar{g}_{\chi}=1$, for each $l$ and $i$
 Eq.~(\ref{md uni})  is satisfied if
\begin{eqnarray}
\sqrt{s} &\lesssim& 2.5 \ {\rm TeV} \,.
\end{eqnarray}
\\



\begin{thebibliography}{99}

 \bibitem{ivd}
 J.~L.~Feng, J.~Kumar, D.~Marfatia and D.~Sanford,
  Phys.\ Lett.\ B {\bf 703}, 124 (2011)
  [arXiv:1102.4331 [hep-ph]].
 
 
 \bibitem{dama}
 R.~Bernabei {\it et al.}  [DAMA Collaboration],
  Eur.\ Phys.\ J.\ C {\bf 56}, 333 (2008)
  [arXiv:0804.2741 [astro-ph]].
 
 \bibitem{cogent}
  C.~E.~Aalseth {\it et al.}  [CoGeNT Collaboration],
  Phys.\ Rev.\ Lett.\  {\bf 106}, 131301 (2011)
  [arXiv:1002.4703 [astro-ph.CO]].
 
 \bibitem{xenon}
 E.~Aprile, {\it et al.}  [XENON100 Collaboration],
  arXiv:1207.5988 [astro-ph.CO];
  Phys.\ Rev.\ Lett.\  {\bf 107}, 131302 (2011)
  [arXiv:1104.2549 [astro-ph.CO]];
J.~Angle {\it et al.}  [XENON10 Collaboration],
  Phys.\ Rev.\ Lett.\  {\bf 107}, 051301 (2011)
  [arXiv:1104.3088 [astro-ph.CO]].
 
 \bibitem{cdms}
 D.~S.~Akerib {\it et al.}  [CDMS Collaboration],
  Phys.\ Rev.\ D {\bf 82}, 122004 (2010)
  [arXiv:1010.4290 [astro-ph.CO]];
 Z.~Ahmed {\it et al.}  [CDMS-II Collaboration],
  Phys.\ Rev.\ Lett.\  {\bf 106}, 131302 (2011)
  [arXiv:1011.2482 [astro-ph.CO]].
  
 \bibitem{lux}
 D.~S.~Akerib {\it at al.},
 Phys.\ Rev.\ Lett.\  {\bf 112}, 091303(2014)
 [arXiv:1310.8214 [astro-ph.CO]].
 
 \bibitem{gresham} 
  M.~I.~Gresham and K.~M.~Zurek,
  Phys.\ Rev.\ D {\bf 89}, 016017 (2014)
  [arXiv:1311.2082 [hep-ph]].
 
 \bibitem{supercdms}
  R.~Agnese {\it et al.}, 
  arXiv:1402.7137 [hep-ex].
 
 \bibitem{cdex}
  W.~Zhao {\it et al.},
  arXiv:1404.4946 [hep-ex].
 
 \bibitem{mg}
 J.~Alwall, P.~Demin, S.~de Visscher, R.~Frederix, M.~Herquet, F.~Maltoni and T.~Plehn {\it et al.},
  JHEP {\bf 0709}, 028 (2007)
  [arXiv:0706.2334 [hep-ph]];
  J.~Alwall, M.~Herquet, F.~Maltoni, O.~Mattelaer and T.~Stelzer,
  JHEP {\bf 1106}, 128 (2011)
  [arXiv:1106.0522 [hep-ph]].

 \bibitem{pythia}
 T.~Sjostrand, S.~Mrenna and P.~Z.~Skands,
  JHEP {\bf 0605}, 026 (2006)
  [hep-ph/0603175].
 
  \bibitem{pgs}
 J. Conway et al., PGS (Pretty Good Simulation),
 http://physics.ucdavis.edu/\~{ }conway/
 
 research/software/pgs/pgs4-general.htm
 
 \bibitem{fox}
 P.~J.~Fox, R.~Harnik, J.~Kopp and Y.~Tsai,
  Phys.\ Rev.\ D {\bf 85}, 056011 (2012)
  [arXiv:1109.4398 [hep-ph]].
 

 
 \bibitem{atlas}
 G.~Aad {\it et al.}  [ATLAS Collaboration],
  Phys.\ Lett.\ B {\bf 705}, 294 (2011)
  [arXiv:1106.5327 [hep-ex]];
 Tech. Rep. ATLAS-CONF-2011-096, CERN, Geneva (2011).

 \bibitem{mlm}
 M.~L.~Mangano, M.~Moretti and R.~Pittau,
  Nucl.\ Phys.\ B {\bf 632}, 343 (2002)
  [hep-ph/0108069];
M.~L.~Mangano, M.~Moretti, F.~Piccinini and M.~Treccani,
  JHEP {\bf 0701}, 013 (2007)
  [hep-ph/0611129].
 
 \bibitem{cms}
 S.~Chatrchyan {\it et al.}  [CMS Collaboration],
  Phys.\ Rev.\ Lett.\  {\bf 107}, 201804 (2011)
  [arXiv:1106.4775 [hep-ex]];
 Tech. Rep. CMS-PAS-EXO-11-059, CERN, Geneva (2011).
 
 
 \bibitem{cluster MT}
 V.~D.~Barger, A.~D.~Martin and R.~J.~N.~Phillips,
  Phys.\ Lett.\ B {\bf 125}, 339 (1983).
 
 \bibitem{null zone}
 R.~W.~Brown, K.~L.~Kowalski and S.~J.~Brodsky,
  Phys.\ Rev.\ D {\bf 28}, 624 (1983).
 
 
 \bibitem{cms-monoj}
 CMS Collaboration,
 CMS-PAS-EXO-12-048
 (http://cds.cern.ch/record/1525585)
 
 
 \bibitem{cms-monog}
 CMS Collaboration,
 Phys. \ Rev. \ Lett. {\bf 108}, 261803 (2012).
 [arXiv:1204.0821 [hep-ex]].
 
  
 \bibitem{8tev monojet}
 ATLAS Collaboration,
 ATLAS-CONF-2012-147
 (http://cds.cern.ch/record/1493486)
 
 \bibitem{kias}
 KIAS School on MadGraph for LHC Physics Simulation (24-29, October 2011, KIAS, Seoul,
 http://workshop.kias.re.kr/MGLP).
 
 

 

\end{thebibliography}
\end{document}